\documentclass[aps, prb, reprint, amssymb, amsmath, floatfix]{revtex4-2} 

\usepackage[T1]{fontenc}
\usepackage{graphicx}
\usepackage{caption}
\usepackage{subcaption} 
\usepackage{ragged2e}
\usepackage{xparse} 
\usepackage{multirow} 
\usepackage[table]{xcolor} 
\usepackage{booktabs}
\usepackage{tikz}

\usepackage{hyperref}
\hypersetup{
colorlinks = true, linkcolor = [rgb]{0,0,1}, citecolor = [rgb]{0,0,1},
                            urlcolor = [rgb]{0,0,1}}

\newcommand{\de}[1]{\left(#1 \right)}
\newcommand{\De}[1]{\left[#1 \right]}

\newcommand{\mean}[1]{\left<#1 \right>}

\newcommand{\mH}{{\mathcal H}}
\newcommand{\mD}{{\mathcal D}}
\newcommand{\Jone}{{J}_1}
\newcommand{\Jtwo}{{J}_2}

\newcommand{\Hk}{\hat{H}_{\bk}}
\newcommand{\HB}{\hat{H}_{B}}
\newcommand{\Ak}{\hat{A}_{\bk}}
\newcommand{\Bk}{\hat{B}_{\bk}}
\newcommand{\Tk}{\hat{T}_{\bk}}
\newcommand{\Tkd}{\hat{T}_{\bk}^{\dag}}
\newcommand{\Ib}{\hat{I}_{B}}
\newcommand{\Hdiag}{\hat{H}^{\textrm{diag}}_{\mathbf{k}}}
\newcommand{\wk}{\omega_{\mathbf{k}}}

\newcommand{\bS}{{\mathbf S}}
\newcommand{\bD}{{\mathbf D}}
\newcommand{\bb}{{\mathbf b}}   
\newcommand{\bB}{{\mathbf B}}
\newcommand{\bh}{{\mathbf h}}
\newcommand{\bk}{{\mathbf k}}
\newcommand{\bp}{{\mathbf p}}
\newcommand{\bq}{{\mathbf q}}
\newcommand{\bQ}{{\bf Q}}
\newcommand{\bPsi}{{\mathbf \Psi}}
\newcommand{\bPsid}{{\mathbf \Psi}^{\dag}}
\newcommand{\bPhi}{{\mathbf \Phi}}
\newcommand{\bPhid}{{\mathbf \Phi}^{\dag}}

\renewcommand{\sd}{s^{\dag}}
\newcommand{\td}{t^{\dag}}
\newcommand{\bd}{b^{\dag}}

\newcommand{\ket}[1]{\left| #1 \right>}

\newcommand{\ua}{\uparrow}
\newcommand{\da}{\downarrow}

\newcommand{\pp}[2]{\frac{\partial #1}{\partial #2}}

\begin{document}


\title{Triplon thermal Hall response in a honeycomb-lattice $J_1$-$J_2$ Heisenberg antiferromagnet with Dzyaloshinskii--Moriya interactions}
\author{Guilherme F. Figueiredo}
\affiliation{Instituto de Física Gleb Wataghin, Universidade Estadual de Campinas (UNICAMP), 13083-859 Campinas, SP, Brazil} 
\date{\today} 

\author{E. Miranda}
\affiliation{Instituto de Física Gleb Wataghin, Universidade Estadual de Campinas (UNICAMP), 13083-859 Campinas, SP, Brazil}
\date{\today} 

\author{R. L. Doretto}
\affiliation{Instituto de Física Gleb Wataghin, Universidade Estadual de Campinas (UNICAMP), 13083-859 Campinas, SP, Brazil}
\date{\today} 


\begin{abstract}
We investigate triplon topology and the thermal Hall transport in a columnar
valence-bond solid of a spin-$1/2$ honeycomb-lattice $J_1$-$J_2$
Heisenberg antiferromagnet. Inequivalent triangular sublattices, modeled by
$J_2^A\neq J_2^B$, remove inversion symmetry at nearest-neighbor bond centers
and allow in-plane Dzyaloshinskii--Moriya (DM) interactions.  We identify the
two symmetry-allowed in-plane DM vector patterns and present detailed 
results for the pattern parallel to the bonds including an out-of-plane magnetic
field.  Within a harmonic bond-operator treatment, we determine the triplon bands, 
their Berry curvatures and Chern numbers, and the triplon thermal Hall conductivity.  
The DM interaction and magnetic field generate nonzero Berry curvatures, but the
numerically integrated Chern number of every isolated triplon band vanishes.
Nevertheless, the energy-dependent Bose weighting of positive and negative
Berry-curvature regions produces a finite thermal Hall response at nonzero
low temperatures.  We delimit the dilute-triplon regime in which the harmonic
transport calculation is expected to be reliable.  These results show that a vanishing Chern number band 
does not directly constrain the triplon response of a valence-bond solid.
We also comment on the effects on the columnar valence-bond solid phase of a DM interaction 
between next-nearest-neighbor spins characterized by an out-of-plane vector pattern.

\end{abstract}


\maketitle

\section{Introduction}
\label{sec:introduction}

Two-dimensional quantum antiferromagnets (AFMs) provide a
setting in which frustration and quantum fluctuations can suppress magnetic
long-range order and stabilize quantum-paramagnetic ground states
\cite{review-frustrated-01,review-frustrated-02}.  These include valence-bond
solids (VBSs), which break lattice symmetries
\cite{review-sachdev,review-sachdev-01}, and quantum spin liquids, which do not
\cite{review-balents-01,review-balents}.
The square- and honeycomb-lattice $J_1$--$J_2$ models are paradigmatic
frustrated AFMs since here one can tune the system through several phases by varying the ratio $J_2/J_1$ 
between the exchange couplings \cite{review-frustrated-01}.
The Hamiltonian of the spin-$1/2$ $J_1$-$J_2$ AFM Heisenberg model is given by
\begin{equation}
\label{eq:j1j2}
    \mH = J_1 \sum_{\mean{ij}} \bS_i \cdot \bS_j + J_2\sum_{\mean{\mean{ij}}}\bS_i \cdot \bS_j,
\end{equation}
where $\bS_i$ is a vector of spin-$1/2$ operators at site $i$ and the parameters $J_1 > 0$ and $J_2 > 0$ are, 
respectively, the nearest-neighbor and next-nearest-neighbor exchange couplings.

\begin{figure*}[t]
    \begin{subfigure}[t]{0.24\textwidth}
        \centering
        \llap{\textbf{(a)}\hspace{-0.5em}}
        \vspace*{-3.0em}
        \includegraphics[width=\textwidth]{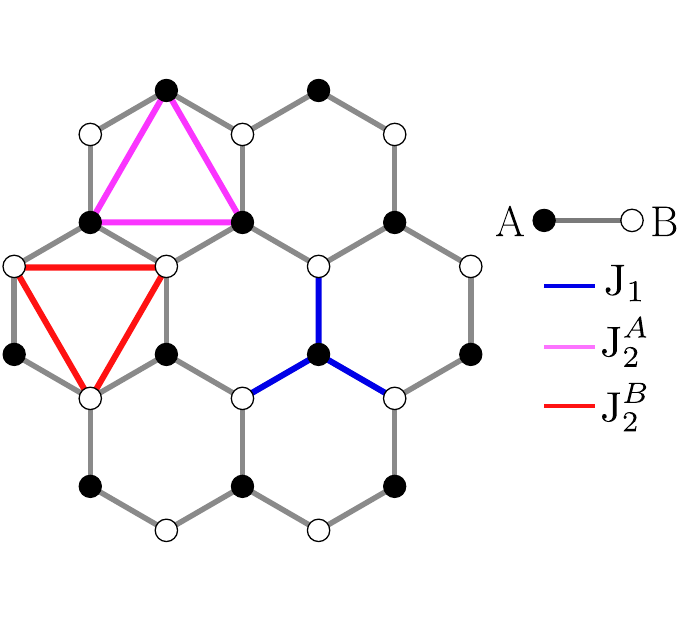}
    \end{subfigure}
    \hspace{0.3cm}
    \begin{subfigure}[t]{0.24\textwidth}
        \centering
        \llap{\textbf{(b)}\hspace{-0.5em}}
        \vspace*{-3.0em}
        \includegraphics[width=\textwidth]{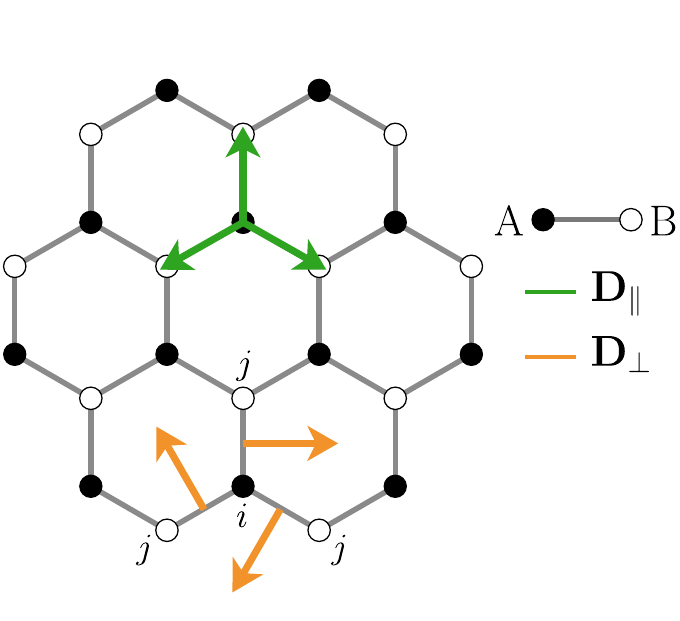}
    \end{subfigure}
    \hspace{0.1cm}
    \begin{subfigure}[t]{0.24\textwidth}
        \centering
        \llap{\textbf{(c)}\hspace{-1.5em}}
        \vspace*{-3.0em}
        \includegraphics[width=\textwidth]{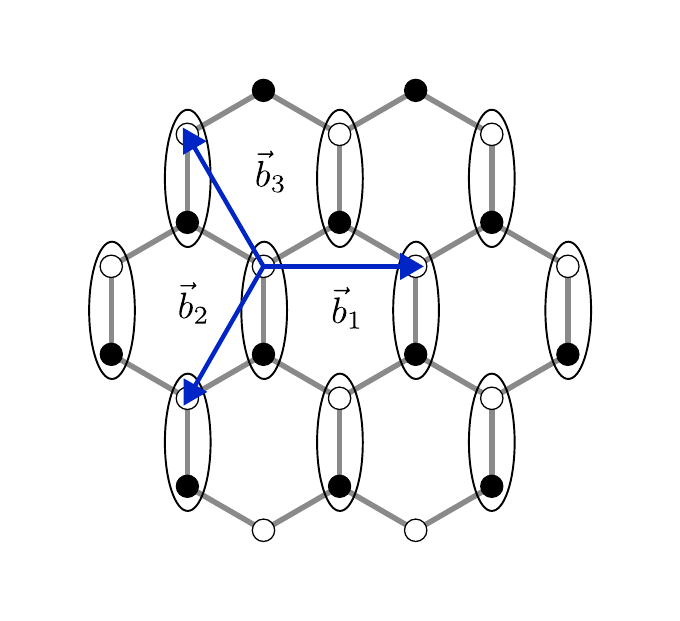}
    \end{subfigure}
    \hspace{0.005cm}
    \begin{subfigure}[t]{0.20\textwidth}
        \centering
        \llap{\textbf{(d)}\hspace{-1.0em}}
        \vspace*{-3.0em}
        \includegraphics[width=\textwidth]{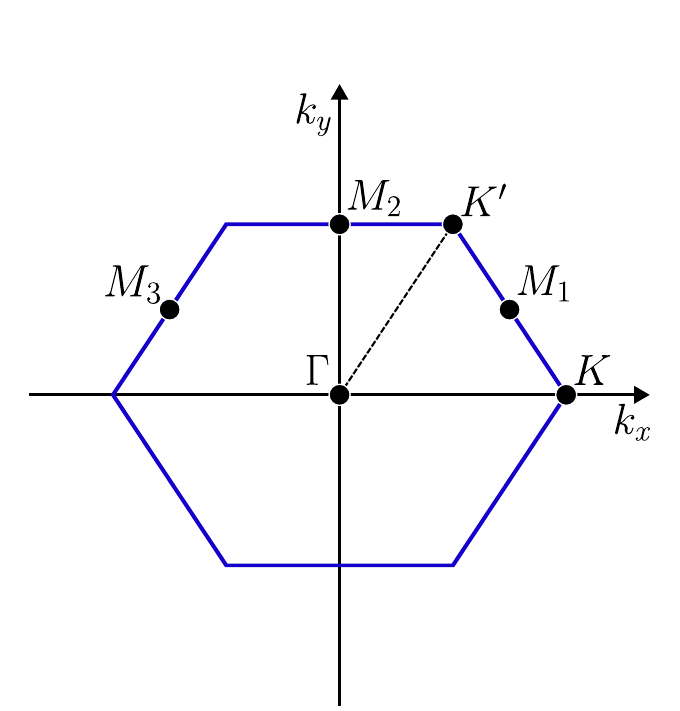}
    \end{subfigure}
    \caption{\justifying{Schematic representations: 
    (a) The spin-$1/2$ $J_1$-$J_2$ AFM Heisenberg model \eqref{eq:ham-j1j2} in the honeycomb lattice, featuring the sublattice-dependent next-nearest-neighbor couplings $J^A_2$ (solid pink line) and $J^B_2$ (solid red line). 
    (b) The DM vectors $\bD_{ij}$ [see Eqs.~\eqref{eq:ham-dm}, \eqref{eq:dim-ham-dm}, and \eqref{eq:d-directions}] allowed by the symmetries of the honeycomb lattice; both parallel $D_\parallel$ and perpendicular $D_\perp$ DM vector patterns are shown. 
    (c) The columnar VBS ground state, where an ellipse indicates that the spins 
    $\bS^1$ (filled circle and associated with the triangular sublattice $A$) and
    $\bS^2$ (open circle and associated with the triangular sublattice $B$) form a singlet state; 
    the vectors $\bb_n$ ($n=1,2,3$) are the dimer nearest-neighbor vectors [Eq.~\eqref{eq:dim-vec}]. 
    (d) The first Brillouin zone of the dimerized lattice $\mD$; high-symmetry points are labeled: 
    $\mathbf{K} = \de{4 \pi / 3\sqrt{3}, 0}$, $\mathbf{K}' = \de{2 \pi / 3 \sqrt{3}, 2 \pi / 3}$, 
    $\mathbf{M}_1 = \de{\pi / \sqrt{3}, \pi / 3}$, $\mathbf{M}_2 = \de{0, 2 \pi / 3}$, and $\mathbf{M}_3=\de{-\pi/\sqrt{3}, \pi/3}$;
    the lattice spacing of the original honeycomb lattice $a = 1$.} }
    \label{fig:model}
\end{figure*}

Several studies have been devoted to the honeycomb lattice $J_1$-$J_2$ AFM Heisenberg model 
\cite{mulder10,albuquerque11,ronny11,singh11,campbell12,zhang13,ganesh2013,
zhu13,fisher13,ciolo14,bieri17,arnaud18,Yu2022}.
In particular, the phase diagram of the model at zero temperature has been determined with the density 
matrix renormalization group (DMRG) \cite{ganesh2013,zhu13}.
The results indicate that, for $J_2 \lesssim 0.24\, J_1$, a   long-range ordered Néel phase, 
with ordering wave vector $\bQ_0 = (0,0)$, sets in. As the ratio $J_2/J_1$ increases,
a plaquette VBS phase is stable for $0.24\, J_1  \lesssim J_2 \lesssim 0.35\, J_1$. 
Finally, a columnar VBS (also referred to as a staggered or nematic VBS in the literature) 
might be the ground state for $0.35\, J_1  \lesssim J_2 \lesssim 0.50\, J_1$.
These results are in agreement with Monte Carlo simulations based on variational fermionic wave functions
\cite{bieri17,becca20}. 
Schwinger boson mean-field approaches \cite{zhang13,arnaud18} also find evidence for a gapped quantum paramagnetic
phase in the intermediate parameter region  $0.21\, J_1  \lesssim J_2 \lesssim 0.40\, J_1$, 
while Néel and spiral ordered phases are stable for lower and larger $J_2/J_1$ ratios, respectively.
It should be recalled that 
the classical ($S\rightarrow\infty$) model of Eq.~\eqref{eq:j1j2} has Néel
order for $J_2/J_1<1/6$ and a continuously degenerate manifold of spiral ground
states for $J_2/J_1>1/6$ \cite{fouet01,baez17,hikaru19,consoli24,rosales25}.
This manifold underlies the regime commonly termed a classical spiral spin
liquid \cite{classical-SL19}: Its ordering wave vectors $\bQ_0$ form closed
contours around
the $\Gamma$ (for $1/6 < J_2/J_1 < 1/2$) 
and  $K$ (for $1/2 < J_2/J_1 < 1$) points of the first Brillouin zone,
see, e.g., Fig.~1 from Ref.~\cite{rosales25}.

%
The effects of an additional Dzyaloshinskii-Moriya (DM) interaction term \cite{dm01,dm03,dm02,camley23,fert23} between localized spins
in a honeycomb lattice have also been considered, 
both for ferromagnets (FMs) \cite{owerre16,owerre16-2,chen18,wu21,kim22,Li26}
and AFMs \cite{okubo12,xiao16,kovalev16,hoon18,joshi19}.
In particular, some attention has been devoted to the fact that a DM interaction may give rise to phases with topologically nontrivial excitations \cite{murakami17,review-top-magnon}, since the DM interaction is related to spin-orbit coupling, and therefore, it could act as a pseudomagnetic field for the (charge neutral) excitations of an insulating quantum magnet.
In long-range-ordered ground states, 
the DM interaction may give rise to magnons with topologically nontrivial  bands \cite{owerre16,owerre16-2,wu21,kim22}, 
while, in VBS phases, one could find topologically nontrivial triplon excitations \cite{penc15,malki17,coldea17,joshi17}.  
In case the topologically nontrivial excitation bands are characterized by finite Chern numbers \cite{review-top-insulator}, 
such an insulating quantum magnet could be seen as a bosonic analogue 
of a Chern band (electronic) insulator \cite{haldane88,review-top-insulator}.

Because magnons and triplons are charge neutral, their band topology
does not produce an electronic charge Hall current.  It can instead be probed
through their longitudinal and transverse (Hall) heat currents in response to
a temperature gradient.
Therefore, the nontrivial topological properties of a magnetic insulator can be probed, e.g., via 
measurements of the thermal Hall conductivity $\kappa_{xy}$ \cite{murakami17,review-zhang}. 
In addition to honeycomb lattice FMs \cite{owerre16,wu21,kim22,zhou26}
and AFMs \cite{mertig22}, a thermal Hall effect due to magnons has been studied in 
FMs on Shastry-Sutherland \cite{malki19},
pyrochlore \cite{tokura12},
and perovskite \cite{tokura12} lattices and
in AFMs on kagome \cite{fiete18} and square \cite{hotta19} lattices.
Moreover, a thermal Hall effect due to triplons has been predicted for the
Shastry-Sutherland compound SrCu$_2$(BO$_3$)$_2$ \cite{penc15,malki17}
(although not experimentally observed \cite{matsuda22})
and for AFMs on square \cite{buzo24} and maple-leaf and star lattices \cite{esaki25}.
A thermal Hall effect also due to triplons, but induced by an electric field, has been proposed  
for the quantum dimer magnets $X$CuCl$_3$ ($X$ = Tl, K) \cite{esaki23}.

In addition to the thermal Hall effect, a magnon spin Nernst effect
(an effect similar to the spin Hall one for electrons \cite{review-SHE})
in honeycomb lattice AFMs \cite{xiao16,kovalev16,nasu25} has also 
received some attention. In contrast to the thermal Hall effect, 
the spin Nernst one might be observed 
in phases that preserve time-reversal symmetry.

%
The topological properties of a columnar VBS in a square-lattice AFM
were recently studied in Ref.~\cite{buzo24}.
Here, the square lattice insulating quantum magnet was described by the spin-$1/2$ $J_1$-$J_2$ AFM
Heisenberg model \eqref{eq:j1j2} with an additional DM interaction between nearest-neighbor spins and an external magnetic 
field. The focus was on the intermediate parameter region of the square lattice $J_1$-$J_2$ model 
($0.4\, J_1 \lesssim J_2 \lesssim 0.6\, J_1$)
with the DM-vector pattern characteristic of the CuO$_2$ planes of
YBa$_2$Cu$_3$O$_6$ (YBCO) \cite{dm-cuprates02}.
The columnar VBS phase was described by an effective interacting boson model written in terms of triplet operators which was determined within the bond-operator representation for the spin operators \cite{sachdev90}.
The interacting boson model was treated within the harmonic
approximation and the triplon excitation bands were determined.
It was found that the DM interaction yields a finite Berry curvature for the
triplon bands, but the corresponding Chern numbers vanish, i.e., such a VBS phase is characterized by topologically trivial triplon excitations. 
Moreover, the thermal Hall conductivity $\kappa_{xy}$ as a function of the temperature due to triplon excitations was obtained with the formalism that was  employed in the study of the thermal Hall effect due to magnons \cite{prl-ryo11,prb-ryo11}.
In spite of the fact that the triplon excitations are topologically trivial, the
thermal Hall conductivity of the columnar VBS phase on the square lattice is finite.
The study in Ref.~\cite{buzo24} was motivated by the results of Katsura {\sl et al.} \cite{lee10}, according to which  the thermal Hall conductivity of an ordered magnet should vanish (a {\sl no-go theorem}) in edge-sharing lattices, such as the honeycomb, square, and triangular lattices. The results of Ref.~\cite{buzo24} indicate that such a restriction does not apply to a square lattice VBS phase.

%
Here we extend the procedure of Ref.~\cite{buzo24} to a honeycomb
lattice.  We consider a spin-$1/2$ $J_1$--$J_2$ AFM Heisenberg model with inequivalent
next-nearest-neighbor couplings, $J_2^A\neq J_2^B$ [Fig.~\ref{fig:model}(a)], a
nearest-neighbor DM interaction [Fig.~\ref{fig:model}(b)], and an external
magnetic field.  We focus on the columnar VBS illustrated in Fig.~\ref{fig:model}(c) and
calculate the thermal Hall conductivity $\kappa_{xy}$ of its triplon
excitations.
Our central aims are to verify (a) whether a DM interaction yields topologically 
nontrivial triplon bands and (b) whether a VBS phase in an edge-sharing honeycomb 
lattice also displays a finite thermal Hall conductivity as found for 
the square-lattice AFM \cite{buzo24}. 
Our answers are ``no'' to (a) and ``yes'' to (b). The latter confirms that 
the {\sl no-go theorem} \cite{lee10} proposed for ordered magnets does not apply to VBS phases.

Our paper is organized as follows.
In Sec.~\ref{sec:model}, we introduce the Hamiltonian of the frustrated spin-$1/2$ 
AFM Heisenberg model on the honeycomb lattice, including the details of the DM interaction. 
We briefly review the bond-operator formalism \cite{sachdev90} employed in the 
description of the columnar VBS phase in Sec.~\ref{sec:bond-operator-formalism},
while the effective interacting boson model in terms of triplet operators, 
determined within the bond-operator formalism, is shown 
in Sec.~\ref{sec:effective-boson-model}.
In Sec.~\ref{sec:harmonic-approximation}, we consider the effective boson model 
within the harmonic approximation and, in particular, determine the region of stability of
the columnar VBS phase and the triplon excitation bands. Only results for the 
parallel DM vector pattern are discussed. Some details of the (numerical) procedure
employed to diagonalize the effective boson model are also provided. 
The Berry curvatures of the triplon excitation bands and the respective Chern numbers
are calculated in Sec.~\ref{sec:berry-curvatures-and-chern-numbers},
while the thermal Hall conductivity $\kappa_{xy}$ due to the triplons is obtained in Sec.~\ref{sec:thermal-hall-conductivity}.
Finally, in Sec.~\ref{sec:summary-and-conlusions}, we summarize our main findings.  
Further details concerning the effective boson model in real space and
a brief discussion of the VBS phase in the presence of a DM interaction between
next-nearest-neighbors, with an out-of-plane DM vector pattern,
can be found in the two Appendices.

\section{MODEL}
\label{sec:model}

We consider a frustrated spin-$1/2$ Heisenberg AFM on a honeycomb lattice described by the Hamiltonian
\begin{equation}
    \label{eq:ham}
    \mathcal{H} = \mH_J + \mH_{DM} + \mH_{B}.
\end{equation}
The first term $\mH_J$ is a modified $J_1$-$J_2$ model [see Eq.~\eqref{eq:j1j2}], namely, 
\begin{equation}
    \label{eq:ham-j1j2}
      \mH_J = J_1 \sum_{\mean{ij}} \bS_i \cdot \bS_j + \sum_{\mean{\mean{ij}}} J^a_2\,\bS_i \cdot \bS_j,
\end{equation}
where $\bS_i$ is a vector spin-$1/2$ operator at site $i$ and the parameters $J_1 > 0$ and $J^a_2 > 0$, with $a = A, B$, are the nearest-neighbor and next-nearest-neighbor exchange couplings, respectively [see Fig.~\ref{fig:model}(a)].
Note that we take the latter to be different on the two triangular sublattices $A$ and $B$, i.e., we assume different atoms occupy the  $A$ and $B$ sublattices.
The relevance of this parameter choice for our study is discussed in detail in Sec.~\ref{sec:effective-boson-model}.
The $\mH_{DM}$ term is a Dzyaloshinskii-Moriya interaction \cite{dm01,dm02,dm03,camley23,fert23} between nearest-neighbor spins,
\begin{equation}
    \label{eq:ham-dm}
    \mH_{DM} = \sum_{\mean{ij}} \bD_{ij} \cdot \de{\bS_i \times \bS_j},
\end{equation}
where $\bD_{ij}$ is a DM vector that couples the spins at sites $i$ and $j$;
we display in Fig.~\ref{fig:model}(b) the two in-plane DM patterns allowed by the symmetries of a
honeycomb lattice with two inequivalent sites per unit cell \cite{dm02}: a
pattern $\bD^\parallel_{ij}$ parallel to the nearest-neighbor bonds and a
perpendicular pattern $\bD^\perp_{ij}$.  
A detailed analysis for the parallel pattern is presented below.
Finally, $\mH_B$ is a Zeeman term, which couples a spin $\bS_i$ with an external magnetic field, 
\begin{equation}
    \label{eq:ham-b}
    \mH_{B} = -g \mu_B \bB \cdot \sum_{i} \bS_i \equiv -\bh \cdot \sum_{i} \bS_i,
\end{equation}
where $g$ is the electron gyromagnetic ratio and $\mu_B$ is the Bohr magneton.

In the following, we concentrate on the columnar VBS phase [Fig.~\ref{fig:model}(c)] of the model of Eq.~\eqref{eq:ham}.

\section{Bond operator formalism}
\label{sec:bond-operator-formalism}

The columnar VBS phase of the AFM Heisenberg model~\eqref{eq:ham} on a honeycomb lattice can be described within the bond-operator representation for spin operators proposed by Sachdev and Bhatt \cite{sachdev90}. In this section, we briefly review the bond-operator representation following the lines of Refs.~\cite{wang97,leite19,doretto20,buzo24}.

We start considering the Hilbert space of two $S=1/2$ spins, $\bS^1$ and $\bS^2$, which consists of a singlet state $\ket{s}$ and three triplet states $\ket{t_x}$, $\ket{t_y}$, and $\ket{t_z}$. One can define a set of creation operators $\sd$ and $\td_{\alpha}$, with $\alpha = x,y,z$, which create these four states out of a  fictitious vacuum $\ket{0}$, namely,
\begin{align}
\label{eq:s-t-states}
    \ket{s} &= \sd \ket{0} = \frac{1}{\sqrt{2}} \de{\ket{\ua \da} - \ket{\da \ua}},  \nonumber \\
    \ket{t_x} &= \td_x \ket{0} = \frac{1}{\sqrt{2}} \de{\ket{\da \da} - \ket{\ua \ua}}, \nonumber \\
    \ket{t_y} &= \td_y \ket{0} = \frac{i}{\sqrt{2}} \de{\ket{\da \da} + \ket{\ua \ua}},  \nonumber \\
    \ket{t_z} &= \td_z \ket{0} = \frac{1}{\sqrt{2}} \de{\ket{\ua \da} + \ket{\da \ua}}. 
\end{align}

To ensure that unphysical states are not present in the new (bosonic) Hilbert space, we must introduce the following constraint
\begin{equation}
    \label{eq:constraint}
    \sd s + \sum_{\alpha} \td_{\alpha} t_{\alpha} = 1.
\end{equation}
Based on Eqs.~\eqref{eq:s-t-states} and \eqref{eq:constraint}, and assuming that the operators $s$ and $t_\alpha$ are a set of four independent bosonic operators, it can be shown that the spin operators $\bS^1$ and $\bS^2$ can be expressed as \cite{sachdev90}
\begin{equation}
    \label{eq:bond-op-representation}
    S^{1,2}_{\alpha} = \pm \frac{1}{2} \de{ \sd t_{\alpha} + \td_{\alpha} s \mp i \epsilon_{\alpha \beta \gamma} \td_{\beta} t_{\gamma} },
\end{equation}
where the indices $\alpha$, $\beta$ and $\gamma$ take the values $x$, $y$ and $z$, $\epsilon_{\alpha \beta \gamma}$ is the totally antisymmetric tensor with $\epsilon_{xyz} = 1$, and the summation convention over repeated indices is used. 
The validity of the representation in Eq.~\eqref{eq:bond-op-representation} can be confirmed by evaluating the matrix elements between the $\sd \ket{0}$ and $\td_\alpha\ket{0}$ states 
[considering the constraint of Eq.~\eqref{eq:constraint}]   
and comparing them with the known matrix elements of the spin operators $S^{1,2}_{\alpha}$ determined within the states $\ket{s}$ and $\ket{t_\alpha}$. 
Also, the SU(2) algebra of the spin operators can be reproduced by the representation of Eq.~\eqref{eq:bond-op-representation}, 
as shown in Ref.~\cite{sachdev90}.

The bond-operator representation of Eq.~\eqref{eq:bond-op-representation} can be easily generalized to any spin operator $\bS_i$ defined at a site $i$ of a given lattice after including a site index $i$ in the singlet $s_i$ and triplet $t_{i \alpha}$ boson operators.


\section{EFFECTIVE BOSON MODEL}
\label{sec:effective-boson-model}

\subsection{$J_1$-$J_2$ AFM Heisenberg model}
\label{subsec:afm-heisenberg-hamiltonian}

In this section, we employ the bond-operator formalism to map the AFM Heisenberg model of Eq.~\eqref{eq:ham} into an effective model expressed in terms of the bosonic operators $s_i$ and $t_{i \alpha}$. We should note that the procedure adopted in this section was previously employed by one of us in studies regarding the columnar \cite{leite19, doretto20, buzo24} and plaquette \cite{doretto14} VBS phases of Heisenberg AFMs on a square lattice. 

We start expressing the Hamiltonian of Eq.~\eqref{eq:ham} in terms of the underlying (triangular) dimerized lattice $\mD$, which is defined by the dimer (singlet) arrangement corresponding to the columnar VBS phase, as illustrated in Fig.~\ref{fig:model}(c). In particular, the Hamiltonian in Eq.~\eqref{eq:ham-j1j2} of the $J_1$-$J_2$ model assumes the form 
\begin{align}
    \label{eq:dim-ham-j1j2}
    \mH_J = &\sum_{i \in \mD} \bigg[ J_1 \de{ \bS^{1}_{i} \cdot \bS^{2}_{i} + \bS^{1}_{i} \cdot \bS^{2}_{i + 2} +  \bS^{1}_{i} \cdot \bS^{2}_{i - 3} }
    \nonumber \\
    & + J^A_2 \de{ \bS^{1}_{i} \cdot \bS_{i + 1}^{1} + \bS^{1}_{i} \cdot \bS_{i + 2}^{1} + \bS^{1}_{i} \cdot \bS_{i + 3}^{1} }
    \nonumber \\
    &+ J^B_2 \de{ \bS^{2}_{i} \cdot \bS_{i + 1}^{2} + \bS^{2}_{i} \cdot \bS_{i + 2}^{2} + \bS^{2}_{i} \cdot \bS_{i + 3}^{2} } \bigg],
\end{align}
where the index $i$ denotes a site of the dimerized lattice $\mD$, which has two spins per unit cell, $\bS_i^1$ and $\bS_i^2$, and the index $n= 1,\;2,\;3$ in $\bS^{1,2}_{i+n}$ corresponds to the dimer nearest-neighbor vectors $\bb_{n}$ 
[see Fig.~\ref{fig:model}(c)],
\begin{equation}
 \bb_1 = \sqrt{3} \,a\, \hat{x},  \quad\quad
 \bb_{2,3} =  -\frac{\sqrt{3}}{2} \, a\, \hat{x}  \mp \frac{3}{2} \, a \,\hat{y},   
 \label{eq:dim-vec}
\end{equation}
where $a$ is the original honeycomb lattice spacing, hereafter set to unity. 
One should note that, for the columnar VBS phase on a honeycomb lattice, the spins $\bS_i^1$ and $\bS_i^2$ are respectively associated with the triangular sublattices $A$ and $B$, see Figs.~\ref{fig:model}(a) and \ref{fig:model}(c). 
Substituting the generalized bond-operator representation \eqref{eq:bond-op-representation} into the Hamiltonian (\ref{eq:dim-ham-j1j2}) yields
\begin{equation}
\label{eq:simp-ham-j1j2}
\mH_J = \mH_{J, 0} + \mH_{J, 2} + \mH_{J, 3} + \mH_{J, 4}.
\end{equation}
Each $\mH_{J,n}$ term contains $n$ triplet operators. The explicit expressions of the $\mH_{J,n}$ terms can be found in Appendix~\ref{appendix:effective-boson-model-in-real-space}.

As discussed in Sec.~\ref{sec:bond-operator-formalism}, we need to impose the constraint \eqref{eq:constraint} in order to avoid unphysical states in the bosonic Hilbert space. We will treat the constraint \eqref{eq:constraint} on average through a Lagrange multiplier $\mu$, i.e., we add the term
\begin{equation}
\label{eq:constraint-avg}
-\mu \sum_{i} \de{ \sd_i s_i + \td_{i \alpha} t_{i \alpha} - 1 }
\end{equation}
to the Hamiltonian \eqref{eq:simp-ham-j1j2}. Furthermore, 
within the bond-operator representation, one can see the ground state of the columnar VBS phase as a condensate of singlets $s_i$. We then take 
\begin{equation}
\label{eq:condensation}
\sd_i = s_i = \langle \sd_i \rangle = \langle s_i \rangle \rightarrow \sqrt{N_0},
\end{equation}
where $N_0$ is a constant. 
Substituting Eq.~\eqref{eq:condensation} into the Hamiltonian \eqref{eq:simp-ham-j1j2} yields an effective boson model expressed solely in terms of the triplet operators $t_{i \alpha}$.  
The parameters $\mu$ and $N_0$ introduced above are later self-consistently calculated, see Sec.~\ref{subsec:self-consistent-equations} for details. 

We now turn to the representation of the Hamiltonian \eqref{eq:simp-ham-j1j2} in momentum space. This is achieved by performing a Fourier transform on the triplet operators, namely,
\begin{equation}
\label{eq:fourier-transform}
t_{i \alpha} = \frac{1}{\sqrt{N_{\mathcal{D}}}} \sum_{\bk \in \textrm{BZ}} e^{i \bk \cdot \mathbf{R}_i} t_{ \bk \alpha}.
\end{equation}
Here, $N_{\mathcal{D}} = N/2$ is the number of sites of the dimerized lattice $\mD$, $N$ is the number of sites of the original honeycomb lattice, $\mathbf{R}_i$ is a vector of the dimerized lattice, and the sum over $\bk$ runs over the first Brillouin zone of the dimerized lattice $\mD$ [Fig.~\ref{fig:model}(d)]. 
Substituting Eq.~\eqref{eq:fourier-transform} into the Hamiltonian~\eqref{eq:simp-ham-j1j2},
one finds that the $\mH_{J,n}$ terms assume the following form in momentum space
\begin{align}
    \mH_{J,0} =&-\frac{3}{8} N_0 N J_1 - \frac{1}{2} \mu N \de{ N_0 - 1 },
    \label{eq:fourier-ham0-j1j2} \\
    \nonumber \\
    \mH_{J,2} =&\sum_{\bk} \left[ A_{\bk} \td_{\bk \alpha} t_{ \bk \alpha} + \frac{B_{\bk}}{2}\de{ \td_{\bk \alpha} \td_{ -\bk \alpha} + \textrm{h.c.} }\right],
    \label{eq:fourier-ham2-j1j2} \\ 
    \nonumber \\
    \mH_{J,3} =&\frac{1}{2\sqrt{N_{\mathcal{D}}}} \sum_{\bk, \bp} \bigg[ \xi^{J}_{\bk - \bp} \de{ \td_{\bk - \bp, \alpha} \td_{\bp \beta} t_{\bk \gamma} + \textrm{h.c.} }  
    \nonumber \\
    &+ \zeta^{J}_{\bk - \bp} \de{ \td_{\bk - \bp, \alpha} \td_{\bp \beta} t_{\bk \gamma} - \textrm{h.c.} } \bigg],
    \label{eq:fourier-ham3-j1j2} \\
    \nonumber \\
    \mH_{J,4} =&\frac{1}{2N_{\mathcal{D}}} \epsilon_{\alpha \beta \gamma} \epsilon_{\alpha \mu \nu} \sum_{\bk, \bp, \bq} \lambda_{\bk} \de{ \td_{\bp + \bk, \beta} \td_{\bq - \bk, \mu} t_{\bq \nu} t_{\bp \gamma}},
    \label{eq:fourier-ham4-j1j2}
\end{align}
where the coefficients $A_{\bk}$, $B_{\bk}$, $\xi^{J}_{\bk}$, $\zeta^{J}_{\bk}$, and $\lambda_{\bk}$ read
\begin{align}
    A_{\bk} =& \frac{J_1}{4} - \mu + B_{\bk}, 
    \nonumber \\ \nonumber \\
    B_{\bk} =& -\frac{N_0}{2} \Big[ -\de{ J_2^A + J_2^B } \cos{k_{1}}
             \nonumber \\
             &+ \de{ J_1 - J_2^A - J_2^B } \de{ \cos{k_{2}} + \cos{k_{3}} } \Big], 
    \nonumber \\  \nonumber \\
    \xi^{J}_{\bk} =& \sqrt{N_0} J_1 \de{ \sin{k_{2}} - \sin{k_{3}} }, 
    \label{eq:coefs-hj} \\
    \nonumber \\
    \zeta^{J}_{\bk} =& -i \sqrt{N_0} \de{ J_2^A - J_2^B }\de{ \cos{k_{1}} + \cos{k_{2}} + \cos{k_{3}} },
    \nonumber \\  \nonumber \\
    \lambda_{\bk} =& -\frac{1}{2} \Big[ \de{J^A_2 + J^B_2} \cos{k_{1}}
                   \nonumber \\
                   &+ \de{J_1 + J^A_2 + J^B_2} \de{\cos{k_{2}} + \cos{k_{3}}} \Big],
    \nonumber
\end{align}
with $k_{1}$, $k_{2}$, and $k_{3}$ being defined as
\begin{equation}
    \label{eq:cdot-kb}
    k_{1} \equiv \bk \cdot \mathbf{b}_{1}, \quad k_{2} \equiv \bk \cdot \mathbf{b}_{2}, \quad k_{3}\equiv \bk \cdot \mathbf{b}_{3}.
\end{equation}

In the next sections, the same procedure is applied to the Dzyaloshinskii-Moriya term \eqref{eq:ham-dm} and Zeeman one \eqref{eq:ham-b}.

\subsection{Dzyaloshinskii-Moriya term}
\label{subsec:dzyaloshinskii-moriya-effective-hamiltonian}

In terms of the dimerized lattice $\mD$, the Hamiltonian \eqref{eq:ham-dm} reads
\begin{align}
    \label{eq:dim-ham-dm}
    \mH_{DM} = & \sum_{i \in \mD} \, \bigg[ \bD_{i, i} \cdot \de{ \bS^1_i \times \bS^2_i }
     + \bD_{i, i+2} \cdot \de{ \bS^1_i \times \bS^2_{i+2} }
    \nonumber \\
    & + \bD_{i, i-3} \cdot \de{ \bS^1_i \times \bS^2_{i-3} } \bigg],
\end{align}
with the indices $n = 2, 3$ denoting the nearest-neighbor vectors \eqref{eq:dim-vec}. Here the DM vectors $\bD_{i,j}$ are defined according to the parallel (green vectors) and perpendicular (orange vectors) patterns indicated in Fig.~\ref{fig:model}(b), 
\begin{align}
    &\bD_{i,i} = D_{\perp} \de{ 1, \; 0, \; 0 } + D_{\parallel} \de{ 0, \; 1, \; 0 }, 
    \label{eq:d-directions} \\
    \nonumber \\
    &\bD_{i,i+2} = D_{\perp} \de{ -1/2, \, +\sqrt{3}/2, \, 0 } 
     - D_{\parallel} \de{ \sqrt{3}/2, \, 1/2, \, 0 },
    \nonumber \\
    \nonumber \\
    &\bD_{i,i-3} = -D_{\perp} \de{ 1/2, \, \sqrt{3}/2, \, 0 } 
     + D_{\parallel} \de{ \sqrt{3}/2, \, -1/2, \, 0 }.
    \nonumber
\end{align}

After substituting the generalized bond-operator representation \eqref{eq:bond-op-representation} into the Hamiltonian \eqref{eq:dim-ham-dm}, one finds that
\begin{equation}
\label{eq:simp-ham-dm}
\mH_{DM} = \mH_{DM,1} + \mH_{DM,2} + \mH_{DM, 3} + \mH_{DM, 4},
\end{equation}
where, similarly to the Hamiltonian \eqref{eq:simp-ham-j1j2}, each $\mH_{DM,n}$ term contains $n$ triplet operators (see Appendix~\ref{appendix:effective-boson-model-in-real-space} for details). Including the replacement \eqref{eq:condensation} and performing the Fourier transform \eqref{eq:fourier-transform}, one arrives at that the momentum-space representation of the Hamiltonian \eqref{eq:dim-ham-dm}, namely,
\begin{align}
    \mH_{DM,2} =& i \sum_{\bk} \de{ C_{\bk} \epsilon_{y \beta \gamma} + D_{\bk} \epsilon_{x \beta \gamma} }
    \nonumber \\
    &\times  \Big[ \td_{\bk \beta} t_{\bk \gamma} + \frac{1}{2}\de{ t_{ -\bk \beta} t_{\bk \gamma} - \textrm{h.c.} } \Big],
    \label{eq:fourier-ham2-DM} \\ 
    \nonumber \\
    \mH_{DM,3} =& i \frac{1}{2\sqrt{N_{\mathcal{D}}}} \sum_{\bk, \bp} \bigg[ \de{ \epsilon_{x \beta \gamma} \xi^{DM}_{x, \bk - \bp} + \epsilon_{y \beta \gamma} \xi^{DM}_{y, \bk - \bp} } 
    \nonumber \\
    &\times \epsilon_{\gamma \mu \nu} \de{ \td_{\bk - \bp, \beta} \td_{\bp \mu} t_{\bk \nu} - \td_{\bk \nu} t_{\bk - \bp, \beta} t_{\bp \mu} }\bigg],
    \label{eq:fourier-ham3-DM} \\
    \nonumber \\
    \mH_{DM,4} =& \frac{i}{2N_{\mathcal{D}}} \sum_{\bk, \bp, \bq} \bigg[ \de{ \zeta^{DM}_{x, \bk} \epsilon_{x \beta \gamma} + \zeta^{DM}_{y, \bk} \epsilon_{y \beta \gamma} } 
    \nonumber \\
    &\times \epsilon_{\beta \mu \nu} \epsilon_{\gamma \mu' \nu'} \td_{\bp + \bk, \mu} \td_{\bq - \bk, \mu'} t_{\bp \nu} t_{\bq \nu'} \bigg],
    \label{eq:fourier-ham4-DM}
\end{align}
with the coefficients $C_{\bk}$, $D_{\bk}$, $\xi^{DM}_{x, \bk}$, $\xi^{DM}_{y, \bk}$, $\zeta^{DM}_{x, \bk}$, and $\zeta^{DM}_{y, \bk}$ defined as
\begin{align}
    C_{\bk} =& \frac{N_0}{4} \Big[ D_{\parallel} \de{ \sin{k_{2}} - \sin{k_{3}} } 
                         -\sqrt{3} D_{\perp} \de{ \sin{k_{2}} + \sin{k_{3}} } \Big],
    \nonumber \\ \nonumber \\
    D_{\bk} =& \frac{N_0}{4} \Big[ \sqrt{3} D_{\parallel} \de{ \sin{k_{2}} + \sin{k_{3}} }
                                          + D_{\perp} \de{ \sin{k_{2}} - \sin{k_{3}} }  \Big],
    \nonumber\\  
    \nonumber \\
    \xi_{x, \bk}^{DM} =& \frac{\sqrt{N_0}}{2} \Big[ \sqrt{3} D_{\parallel} \de{ \cos{k_{2}} - \cos{k_{3}} }
     \nonumber \\
                                                 &+ D_{\perp} \de{ \cos{k_{2}} + \cos{k_{3}} }  \Big],
    \nonumber \\
    \nonumber \\
    \xi_{y, \bk}^{DM} =& \frac{\sqrt{N_0}}{2} \Big[ D_{\parallel} \de{ \cos{k_{2}} + \cos{k_{3}} } 
    \nonumber \\
                                                &- \sqrt{3} D_{\perp} \de{ \cos{k_{2}} - \cos{k_{3}} } \Big],
    \label{eq:coefs-dm}  \\ 
    \nonumber \\
    \zeta^{DM}_{x, \bk} =& \frac{1}{4} \Big[ \sqrt{3} D_{\parallel} \de{ \sin{k_{2}} + \sin{k_{3}} }
                                                    + D_{\perp} \de{ \sin{k_{2}} -\sin{k_{3}} } \Big],
    \nonumber \\ \nonumber \\
    \zeta^{DM}_{y, \bk} =& \frac{1}{4} \Big[ D_{\parallel} \de{ \sin{k_{2}} - \sin{k_{3}} }
                                   -\sqrt{3} D_{\perp} \de{ \sin{k_{2}} + \sin{k_{3}} } \Big].
    \nonumber       
\end{align}

A few remarks regarding the choice of an AFM Heisenberg model on a honeycomb lattice with distinct next-nearest-neighbor exchange couplings $J^A_2 \not= J^B_2$ are in order here.
If one considers a honeycomb lattice with equivalent sites $A$ and $B$ associated with the two triangular sublattices (i.e., in the absence of sublattice asymmetry), the inversion symmetry with respect to the midpoint of each nearest-neighbor bond prohibits a DM interaction between nearest-neighbor spins according to Moriya's rules \cite{dm02}. On the other hand, DM interactions between next-nearest-neighbor spins are allowed and, in particular, one finds that both 
in-plane \cite{kim22} and out-of-plane \cite{kim22,wu21,xiao16,Li26}
DM vector patterns have been considered in studies of honeycomb-lattice magnets. 
For a next-nearest-neighbor out-of-plane DM interaction on equivalent
sublattices, $D_z^A=D_z^B$, the quadratic $\mH_{DM,2}$ and quartic
$\mH_{DM,4}$ terms vanish, leaving the cubic interaction as the leading term
(see Appendix~\ref{appendix:DM-nnn} for details).  
Its effects therefore cannot be captured by the harmonic treatment used in the main text.
We therefore consider inequivalent $A$ and $B$ (triangular) sublattices,
phenomenologically encoded by the next-nearest-neighbor exchange couplings $J_2^A\neq J_2^B$ [Eq.~\eqref{eq:ham}].  
This choice removes inversion symmetry at the midpoint of a nearest-neighbor bond,
thereby allowing the nearest-neighbor DM patterns shown in
Fig.~\ref{fig:model}(b). Both the quadratic $\mH_{DM,2}$ and quartic
$\mH_{DM,4}$ terms in Eq.~\eqref{eq:simp-ham-dm} are then finite.

The presence of a linear term $\mathcal{H}_{DM,1}$ in the effective boson model \eqref{eq:simp-ham-dm} also requires a few words, since it couples the singlet and triplet sectors. The procedure to treat the linear term is discussed in detail in Refs.~\cite{coldea17, buzo24}, and therefore, in the following, we briefly summarize the main aspects of the procedure. We focus on the $D_{\parallel}$ vector pattern, i.e., we set the parameter $D_{\perp} = 0$ in Eqs.~\eqref{eq:dim-ham-dm}-\eqref{eq:simp-ham-dm}. 
Since the linear term couples the singlet $s_i$ and triplet $t_{i y}$ operators, in addition to the linear term $\mathcal{H}_{DM,1}$, we should also consider the local terms related to $\mH_{J,0}$ and  $\mH_{J,2}$ [see Eq.~\eqref{eq:real-hamj1j2}], namely,
\begin{equation}
    \label{eq:ham-loc}
     \mH_{L} = J_1 \sum_{i \in \mD} \Big[  -\frac{3}{4} \sd_i s_i + \frac{1}{4} \td_{iy} t_{iy} + i\frac{D_{\parallel}}{2 J_1} \de{ \sd_i t_{iy} - \textrm{h.c.} } \Big].
\end{equation}
In order to remove the linear term, one performs an (exact) unitary transformation, followed by an expansion of the matrix elements of the related $2 \times 2$ unitary matrix up to first order in the parameter $D_{\parallel}/(2 J_1)$.  
One of the effects of such a unitary transformation is to modify the coefficient $C_{\bk}$ [see Eq.~\eqref{eq:coefs-dm}] by adding an extra term $C'_{\bk}$, which is given by
\begin{equation}
    C'_{\bk} = N_0 J_1 \left(\frac{D_\parallel}{2 J_1} \right) \big( \sin{k_{2}} - \sin{k_{3}} \big).
\label{eq:coef-cprime}
\end{equation}
In other words, after removing the linear term, the quadratic term $\mathcal{H}_{DM,2}$ of the effective boson model \eqref{eq:simp-ham-dm} preserves its form [up to first order in the parameter $D_{\parallel}/(2 J_1)$], 
but now the coefficient $C_{\bk} \rightarrow C_{\bk} + C'_{\bk}$, i.e., 
\begin{equation}
    C_\bk = \frac{1}{4}( 1 + 2\eta) N_0 D_\parallel \big( \sin{k_{2}} - \sin{k_{3}} \big).
\label{eq:coef-C2}
\end{equation}
Note that we added the coefficient $\eta = 0, 1$ in Eq.~\eqref{eq:coef-C2}, since, in the following sections, we discuss the features of the columnar VBS both in the absence ($\eta = 0$) and presence ($\eta = 1$) of the corrections due to the linear $\mathcal{H}_{DM,1}$.
The cubic [Eq.~\eqref{eq:fourier-ham3-DM}] and quartic [Eq.~\eqref{eq:fourier-ham4-DM}] terms are also modified, once the corrections due to the linear term are considered; however, since these terms are neglected in the following analysis, the full expressions of the (corrected) cubic and quartic terms are not presented here.
A similar correction for the coefficient $D_{\bk}$ [see Eq.~\eqref{eq:coefs-dm}]  
can be derived when one considers the $D_\perp$ vector pattern.

\subsection{Zeeman term}

The Zeeman term of Eq.~\eqref{eq:ham-b} is written in the dimerized lattice as
\begin{equation}
    \label{eq:dim-hamb}
    \mH_{B} = - h_{\alpha} \sum_{i \in \mD} \de{ S_{i \alpha}^{1} + S_{i \alpha}^{2} },
\end{equation}
where $h_{\alpha} = g \mu_{B} B_{\alpha}$ and the subindex $\alpha = x,\;y,\;z$, denotes the $\alpha$-component of the external magnetic field $\bh  = g \mu_{B} \bB$.

Again, we first expand the Hamiltonian \eqref{eq:dim-hamb} in terms of triplet operators with the aid of the bond-operator representation \eqref{eq:bond-op-representation}. The resulting real-space expression is given by Eq.~\eqref{eq:real-hamb}. We then perform the Fourier transform \eqref{eq:fourier-transform} and obtain the Hamiltonian in momentum space,
\begin{equation}
    \label{eq:fourier-hamb}
    \mH_B = i \epsilon_{\alpha \beta \gamma} \sum_{\bk}  h_{\alpha} \td_{\bk \beta} t_{\bk \gamma}.
\end{equation}


\section{The harmonic approximation}
\label{sec:harmonic-approximation}

Let us now consider the effective boson model for the columnar VBS phase within the lowest-order (harmonic) approximation. 
Within this approximation, we truncate the effective boson model defined by Eqs.~\eqref{eq:simp-ham-j1j2}, \eqref{eq:simp-ham-dm}, 
and \eqref{eq:fourier-hamb} at the quadratic level in triplet operators, thereby ignoring the contributions from cubic 
and quartic triplet-triplet interactions. The effective Hamiltonian then reads
\begin{equation}
    \label{eq:harmonic-approx}
    \mH \approx \mH_{J,0} + \mH_{J,2} + \mH_{DM,2} + \mH_{B}.
\end{equation}
One can express the Hamiltonian of Eq.~\eqref{eq:harmonic-approx} in matrix form as follows
\begin{equation}
    \label{eq:ham-matrix}
    \mH = E_0 + \frac{1}{2} \sum_{\bk} \bPsid_{\bk} \Hk \bPsi_{\bk},
\end{equation}
where $E_0$ is a constant given by
\begin{equation}
    \label{eq:E0}
    E_0 = -\frac{3}{8} N_0 N J_1 - \frac{1}{2} \mu N \de{ N_0 - 1 } - \frac{3}{2} \sum_{\bk} A_{\bk},
\end{equation}
the six-component vector $\bPsid_{\bk}$ is composed of the triplet operators 
\begin{equation}
    \label{eq:psi-dagger}
    \bPsid_{\bk} = 
    \begin{pmatrix}
        \td_{\bk x} & \td_{\bk y} & \td_{\bk z} & t_{-\bk x} & t_{-\bk y} & t_{-\bk z}
    \end{pmatrix},
\end{equation}
and $\Hk$ is a $6\times6$ matrix expressed in terms of the $3\times3$ matrices $\Ak$, $\Bk$ and $\HB$ as follows
\begin{equation}
    \label{eq:Hk}
    \Hk = 
    \begin{pmatrix}
        \Ak + \HB & \Bk \\
        \Bk       & \Ak - \HB
    \end{pmatrix},
\end{equation}

\begin{equation}
    \label{eq:Ak}
    \Ak = 
    \begin{pmatrix}
        A_{\bk}   & 0        & -iC_{\bk} \\
        0         & A_{\bk}  & iD_{\bk} \\
        iC_{\bk} & -iD_{\bk} & A_{\bk}
    \end{pmatrix},
\end{equation}

\begin{equation}
    \label{eq:Bk}
    \Bk = 
    \begin{pmatrix}
        B_{\bk}   & 0        & -iC_{\bk} \\
        0         & B_{\bk}  & iD_{\bk} \\
        iC_{\bk} & -iD_{\bk} & B_{\bk}
    \end{pmatrix},
\end{equation}

\begin{equation}
    \label{eq:HB}
    \HB = 
    \begin{pmatrix}
        0     & ih_z  & -ih_y \\
        -ih_z  & 0     & ih_x \\
        ih_y & -ih_x & 0
    \end{pmatrix}.
\end{equation}
The expressions for the coefficients $A_{\bk}$ and $B_{\bk}$ can be found in Eq.~\eqref{eq:coefs-hj}, the ones for the coefficients $C_{\bk}$ and $D_{\bk}$ are shown in Eq.~\eqref{eq:coefs-dm}, and $h_{\alpha} = g \mu_B B_{\alpha}$ ($\alpha=x,y,z$). 
From now on, we consider the external magnetic field $\bB$ perpendicular to the plane of the honeycomb lattice, i.e., we set $\bB = B \hat{z}$.
Note that, with this choice, the magnetic field $\bB$ is perpendicular to the DM vectors, both the parallel $D_\parallel$ and perpendicular $D_\perp$ patterns [see Fig.~\ref{fig:model}(b)]. The importance of this configuration for our study is discussed in more detail in Sec.~\ref{sec:summary-and-conlusions}.

\subsection{Diagonalization of the quadratic Hamiltonian}
\label{subsec:diagonalization-of-the-quadratic-hamiltonian}

In order to diagonalize the $6 \times 6$ problem defined by Eqs.~\eqref{eq:ham-matrix}-\eqref{eq:Hk}, we follow the procedure described in Refs.~\cite{colpa78, blaizot}. Indeed, a standard unitary transformation is not suitable to diagonalize the matrix \eqref{eq:Hk}, since the bosonic Hamiltonian \eqref{eq:ham-matrix} contains anomalous terms, such as $t_{-\bk x} t_{\bk x}$ and $t_{-\bk x} t_{\bk z}$. Therefore, the matrix \eqref{eq:Hk} should be diagonalized by a paraunitary transformation. 
Such a transformation is performed with the aid of a matrix $\Tk$
that obeys the condition \cite{colpa78}
\begin{equation}
    \label{eq:parauni-prop}
    \Tk \Ib \Tk^{\dag} = \Ib, 
\end{equation}
with the $6 \times 6$ matrix $\Ib$ defined as
\begin{equation}
    \label{eq:Ib}
    \Ib = 
    \begin{pmatrix}
        \hat{I} & 0 \\
        0 & -\hat{I}
    \end{pmatrix},
\end{equation}
where $\hat{I}$ is the $3 \times 3$ identity matrix.

After the diagonalization, the Hamiltonian~\eqref{eq:ham-matrix} reads
\begin{equation}
    \label{eq:diag-ham}
    \mH = E_{GS} + \frac{1}{2} \sum_{\bk}  \bPhid_{\bk} \Hdiag \bPhi_{\bk},
\end{equation}
where
\begin{equation}
    \label{eq:EGS}
    E_{GS} = -\frac{3}{8}  N_0 N J_1 - \frac{1}{2} \mu N \de{ N_0 -1 } + \frac{1}{2} \sum_{\bk, \alpha} \de{ \wk^{\alpha} - A_{\bk} }
\end{equation}
is the VBS ground-state energy,
\begin{equation}
    \label{eq:phi-dagger}
    \bPhid_{\bk} = 
    \begin{pmatrix}
        \bd_{\bk x} & \bd_{\bk y} & \bd_{\bk z} & b_{-\bk x} & b_{-\bk y} & b_{-\bk z}
    \end{pmatrix},
\end{equation}
is a six-component vector whose elements are the boson (triplon) operators $b_{\bk \alpha}$,
and the diagonal $6 \times 6$ matrix $\Hdiag = \Tkd \Hk \Tk$ assumes the form
\begin{equation}
    \label{eq:Hkdiag}
    \Hdiag = \textrm{diag}\de{ \wk^{x},\; \wk^{y},\; \wk^{z},\; \wk^{x},\; \wk^{y},\; \wk^{z} },
\end{equation}
where the eigenvalues $\wk^{\alpha}$ ($\alpha = x,\;y,\;z$) are the energies of the triplon excitations. 
The six-component vectors $\bPhi_\bk$ and $\bPsi_\bk$ 
are related by the matrix $\Tk$ through   
\begin{equation}
    \label{eq:psi-to-phi}
    \bPsi_{\bk} = \Tk \bPhi_{\bk}.
\end{equation}
To determine the transformation matrix $\Tk$, we follow a procedure due to Colpa \cite{colpa78}. Such a procedure is outlined in Sec.~7 of Ref.~\cite{Toth_2015} and briefly summarized below.

The first step to determine the matrix $\Tk$ is to perform a Cholesky decomposition of the matrix $\hat{H}_\bk$ [Eq.~\eqref{eq:Hk}], i.e., we should find a matrix $\hat{K}_\bk$ such that $\hat{H}_\bk = \hat{K}^{\dag}_\bk \hat{K}_\bk$. Next, we solve the eigenvalue problem for the Hermitian matrix $\hat{H}^{\rm aux}_\bk = \hat{K}_\bk \hat{I}_B \hat{K}^{\dag}_\bk$. The eigenvectors are then arranged in a unitary matrix $\hat{U}_\bk$ such that the first 3 diagonal elements of the (diagonal) matrix $\hat{L}_\bk = \hat{U}_\bk^{\dag} \hat{K}_\bk \hat{I}_B \hat{K}_\bk^{\dag} \hat{U}_\bk$ are the positive eigenvalues of $\hat{H}^{\rm aux}_\bk$ while the last 3 diagonal elements are the negative ones. The final expressions for the diagonal Hamiltonian~\eqref{eq:Hkdiag} and the transformation matrix $\Tk$ then read
\begin{align}
    \Hdiag &= \Ib \hat{L}_\bk,
\label{eq:Hdiag-num}  \\
    \Tk &= \hat{K}_\bk^{-1} \hat{U}_\bk \de{\Hdiag}^{1/2}.
\label{eq:Tmatrix}
\end{align}
The procedure described above allows us to numerically determine the transformation matrix $\Tk$. 
A few remarks concerning such a diagonalization procedure are in order here.
We end up with a diagonal matrix \eqref{eq:Hkdiag}, with all eigenvalues $\wk^{\alpha}$ positive, 
only if the Hamiltonian \eqref{eq:Hk} is a positive definite matrix \cite{colpa78}. For that,
the Hamiltonian to be diagonalized must have the form
(Sec.~6 from Ref.~\cite{Toth_2015})
\begin{equation}
    \label{eq:colpa-condition}
    \Hk =
    \begin{bmatrix}
        \hat{a}(\bk) & \hat{b}(\bk) \\
        \hat{b}^\dagger(\bk) & \hat{a}^*(-\bk)
    \end{bmatrix},
\end{equation}
where $\hat{a}(\bk)$ and $\hat{b}(\bk)$ are $n \times n$ complex matrices. We note that such a condition is indeed fulfilled by the Hamiltonian \eqref{eq:Hk}.

Finally, we should mention that an alternative analytical procedure, discussed in Ref.~\cite{buzo24}, 
can also be employed to calculate the matrix $\Tk$.

\subsection{Self-consistency equations}
\label{subsec:self-consistent-equations}

After diagonalizing the Hamiltonian \eqref{eq:harmonic-approx}, one still needs to determine the parameters 
$\mu$ and $N_0$, introduced in Eqs.~\eqref{eq:constraint-avg} and \eqref{eq:condensation}, 
in order to completely determine the ground-state energy~\eqref{eq:EGS} and the triplon dispersion relations $\wk^{\alpha}$ defined in Eq.~\eqref{eq:Hkdiag}.  
Such parameters can be calculated via the saddle-point conditions 
$\partial{E_{GS}}/\partial{\mu}=0$ and $\partial{E_{GS}}/\partial{N_0}=0$, which provide the self-consistency equations
\begin{align}
    \mu =& -\frac{3}{4}\Jone + \frac{1}{2N_{\mathcal{D}}} \sum_{\bk, \alpha} \de{ \pp{\wk^{\alpha}}{N_0} - \pp{B_{\bk}}{N_0} },
    \nonumber \\
    \label{eq:self-const} \\
    N_0 =& 1 + \frac{1}{2N_{\mathcal{D}}} \sum_{\bk, \alpha} \de{ \pp{\wk^{\alpha}}{\mu} + 1 }.
    \nonumber
\end{align}

For definiteness, in the following sections, we will use a fixed ratio between the next-nearest-neighbor exchange couplings $J_2^A$ and $J_2^B$, 
\begin{equation}
\label{eq:j2b}
 J_2^B = 0.9 J_2^A \equiv 0.9 J_2.
\end{equation}
Besides, we will focus on the parallel DM vector pattern $D_{\parallel}$, i.e., we set $D_{\perp} = 0$ in Eq.~\eqref{eq:coefs-dm}. 
Finally, we will assume a fixed external magnetic field $h_z = -0.05\,J_1$.
We will numerically solve the pair of self-consistency equations \eqref{eq:self-const} 
for fixed values of the parameters $J_2$, $D_{\parallel}$, and $h_z$ in order to determine $\mu$ and $N_0$. 

\begin{figure}[t]
    \centering
    \includegraphics[width=0.95\linewidth]{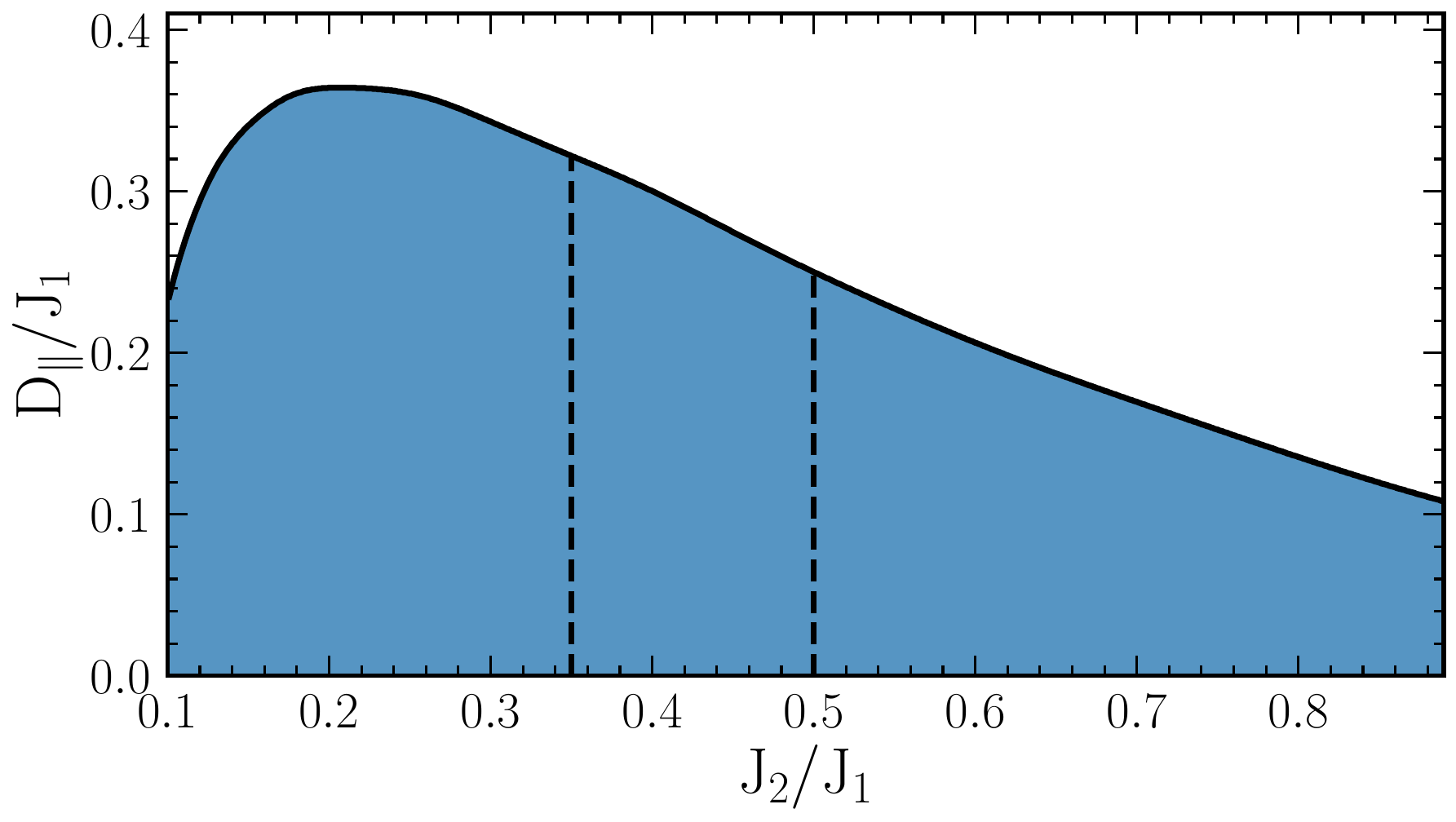}
    \caption{\justifying{Region of stability of the columnar VBS phase [Fig.~\ref{fig:model}(c)] of the Heisenberg model~\eqref{eq:ham} determined within the bond-operator formalism at the harmonic approximation, 
    for a fixed magnetic field $h_z = -0.05 \, J_1$ and neglecting [Eq.~\eqref{eq:coef-C2} with $\eta = 0$]  
    the effects of the linear term \eqref{eq:ham-loc}. 
    The dashed black lines are the boundaries of the VBS phase determined via DMRG for $D_{\parallel} = 0$ and in the absence of the external magnetic field \cite{ganesh2013}.}}
    \label{fig:phase-diagram}
\end{figure}

\subsection{Parallel DM vector pattern: Stability of the columnar VBS phase}
\label{subsec:stability-of-the-vbs}

\begin{figure*}[t]
    \centering
    \begin{subfigure}[b]{0.45\textwidth}
        \centering
        \llap{\textbf{(a)}\hspace{-0.05em}}
        \vspace*{-1.5em}
        \includegraphics[width=\textwidth]{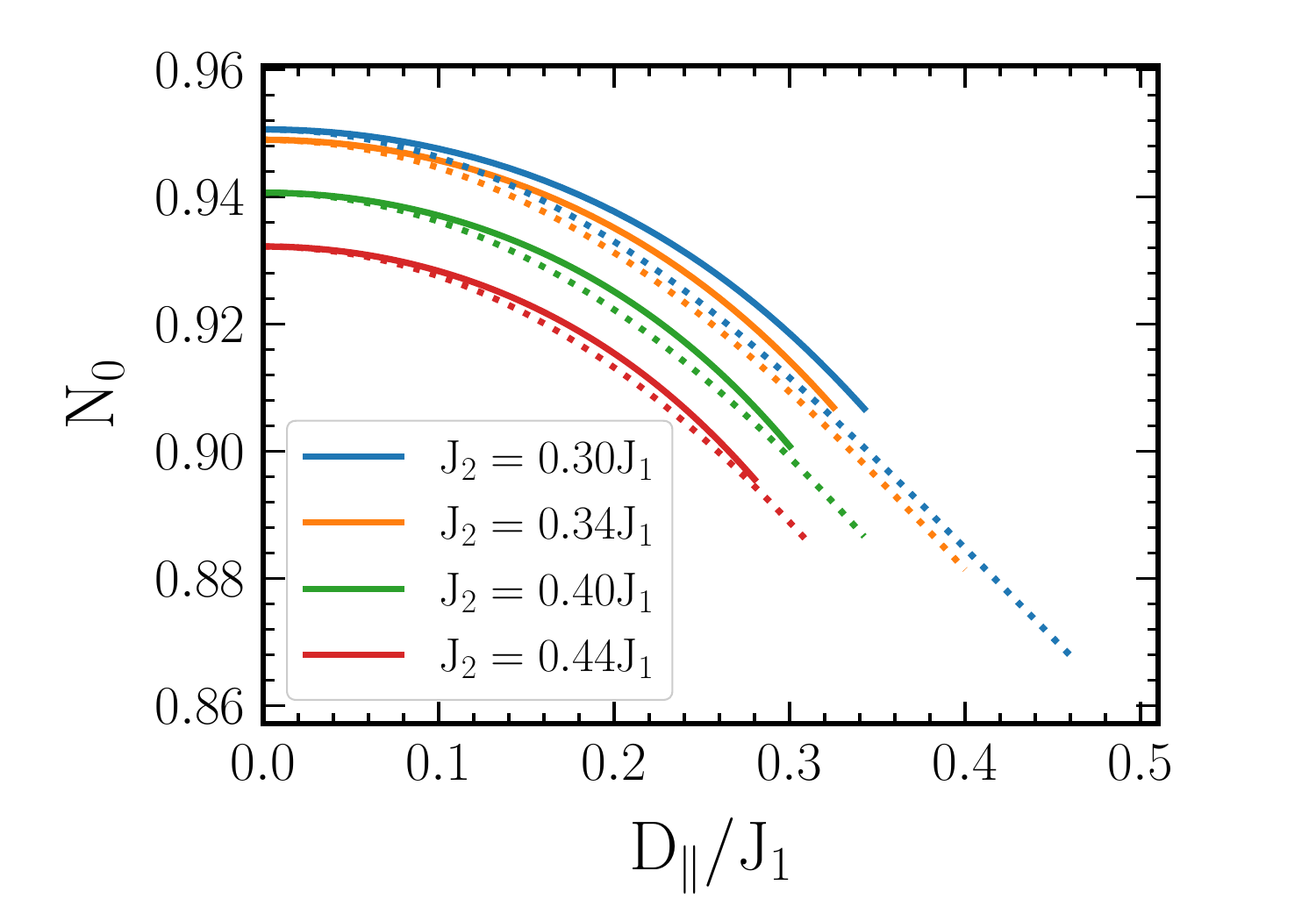}
    \end{subfigure}
    \hspace{0.7cm}
    \begin{subfigure}[b]{0.45\textwidth}
        \centering
        \llap{\textbf{(b)}\hspace{-0.05em}}
        \vspace*{-1.5em}
        \includegraphics[width=\textwidth]{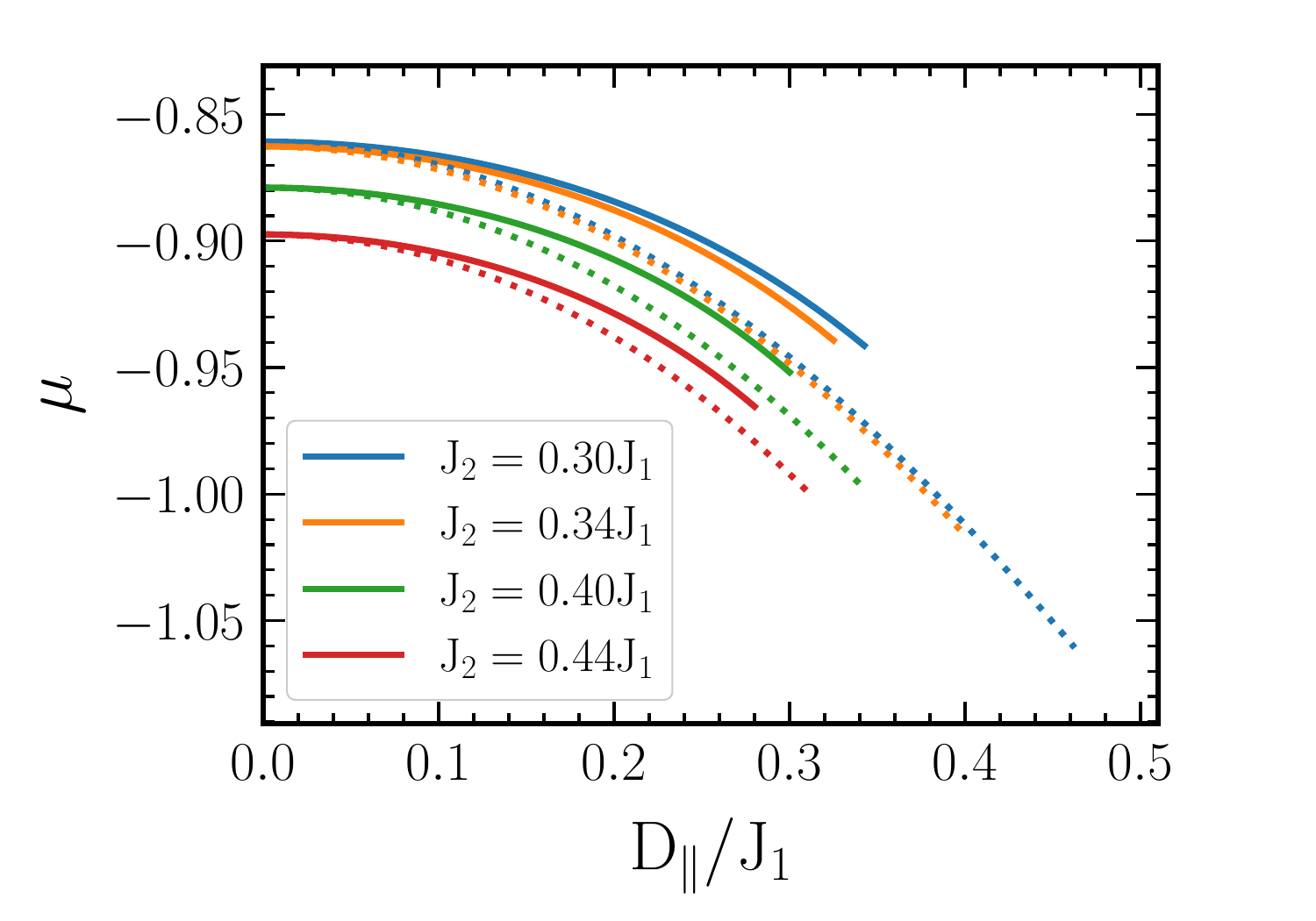}
    \end{subfigure}
    \vspace{0.1cm}
    %
    \begin{subfigure}[b]{0.45\textwidth}
        \centering
        \llap{\textbf{(c)}\hspace{-0.05em}}
        \vspace*{-1.5em}
        \includegraphics[width=\textwidth]{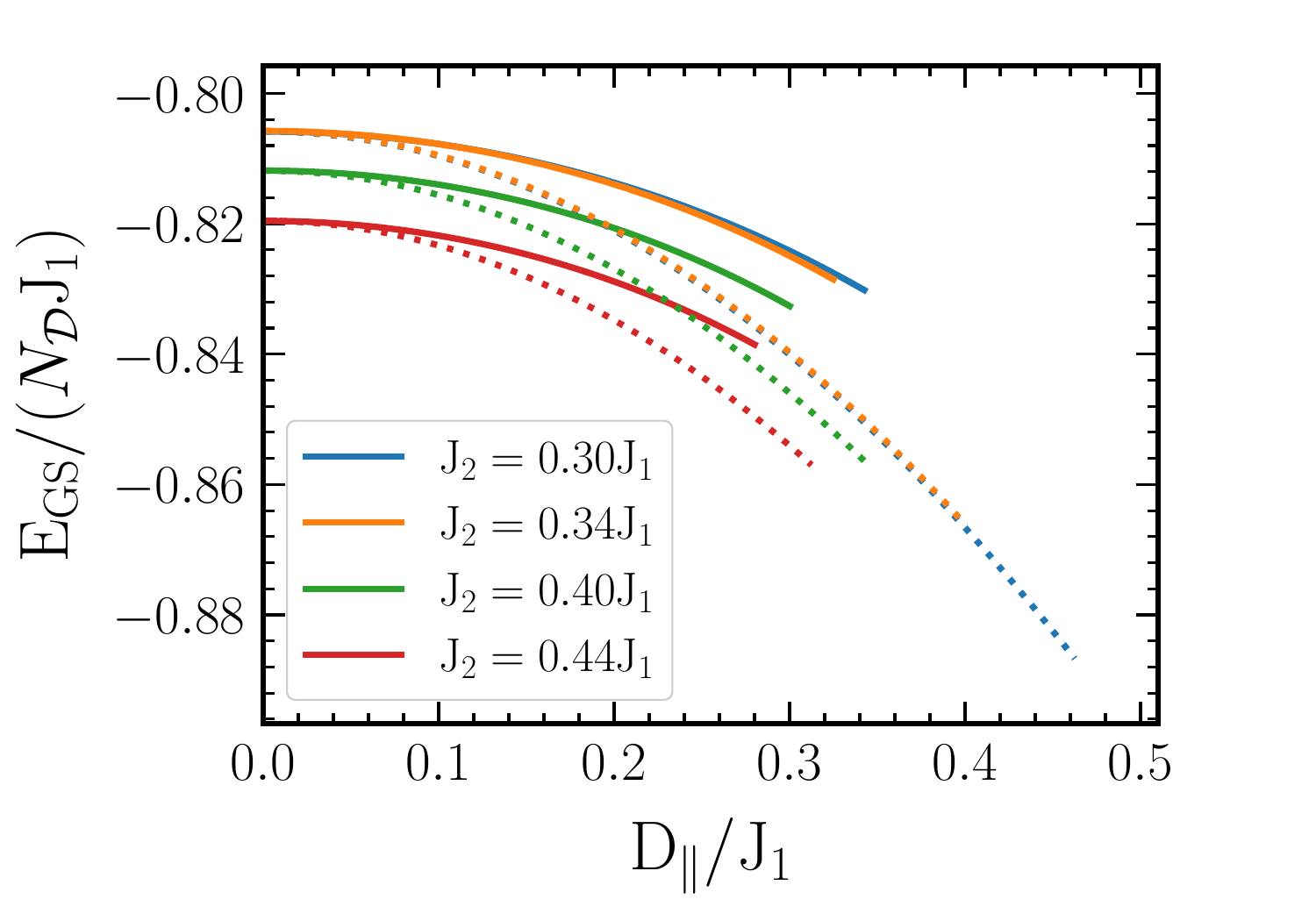}
    \end{subfigure}
    \hspace{0.7cm}
    \begin{subfigure}[b]{0.45\textwidth}
        \centering
        \llap{\textbf{(d)}\hspace{-0.05em}}
        \vspace*{-1.5em}
        \includegraphics[width=\textwidth]{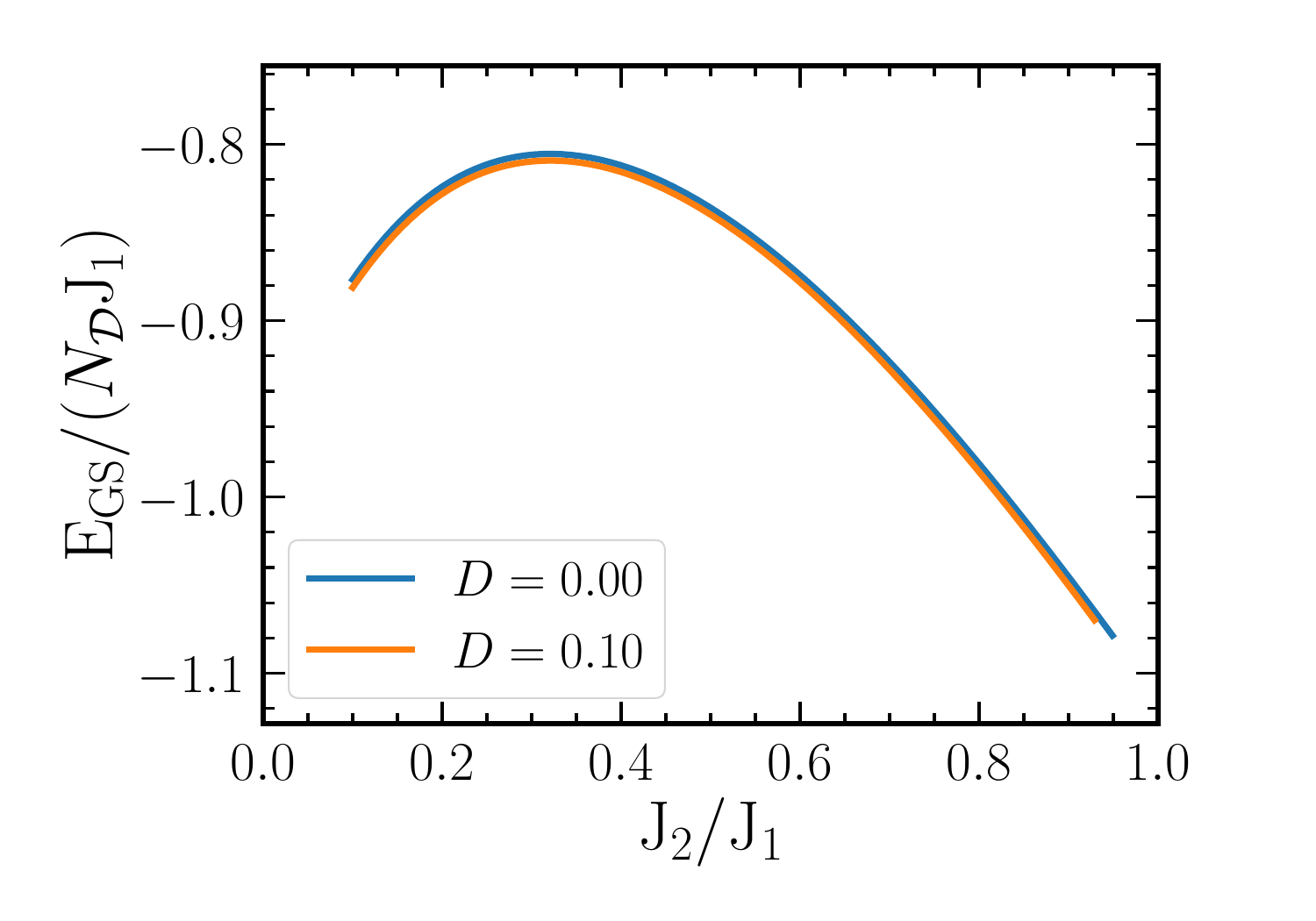}
    \end{subfigure}
    \caption{\justifying{Dependence of several quantities on the DM interaction $D_\parallel/J_1$ 
    for a fixed external magnetic field $h_z = -0.05 \, J_1$:
    The parameters 
    (a) $N_0$ and 
    (b) $\mu$ determined from the self-consistent equations \eqref{eq:self-const}, and  
    (c) the ground-state energy $E_{GS}/(N_{\mathcal{D}} J_1)$.
    Next-nearest-neighbor couplings $J_2^B = 0.9 J_2^A \equiv 0.9 J_2$ [Eq.~\eqref{eq:j2b}], with
    $J_2 = 0.30 \, J_1$ (blue line), $J_2 = 0.34 \, J_1$ (orange line), 
    $J_2 = 0.40 \, J_1$ (green line), and $J_2 = 0.44 \, J_1$ (red line);
    the solid and dotted lines indicate, respectively, results neglecting [Eq.~\eqref{eq:coef-C2} with $\eta = 0$]  
    and including [Eq.~\eqref{eq:coef-C2} with $\eta = 1$] the effects of the linear term \eqref{eq:ham-loc}.
    (d) The ground-state energy $E_{GS}/(N_{\mathcal{D}} J_1)$ [Eq.~\eqref{eq:EGS}] as a function of the next-nearest-neighbor coupling $J_2/J_1$ for DM interaction $D_\parallel = 0$ (solid blue line) and $D_\parallel = 0.10 \, J_1$ (solid orange line).
    $N_{\mathcal{D}} = N/2$ is the number of sites of the dimerized lattice $\mD$, with $N$ being the number of sites of the original honeycomb lattice.
   } }
    \label{fig:params1}
\end{figure*}

The parameter region in which the saddle-point equations \eqref{eq:self-const} have solutions 
and the quadratic Hamiltonian is positive definite is displayed in Fig.~\ref{fig:phase-diagram}.
Here only results in the absence of the corrections due to the linear term \eqref{eq:ham-loc} are shown, see comments below.
We see that, in the absence of the DM interaction, the columnar VBS phase is stable within the parameter region 
$0.10\, J_1 \leq J_2 \leq 0.90 \, J_1$, which is larger than the paramagnetic region 
($0.24\, J_1  \lesssim J_2 \lesssim 0.50\, J_1$) determined by DMRG simulations \cite{ganesh2013,zhu13} 
and the one ($0.21\, J_1  \lesssim J_2 \lesssim 0.40\, J_1$) found within a Schwinger boson mean-field theory formalism \cite{zhang13,arnaud18}. 
We should recall that the DMRG results indicate that the columnar VBS might be the ground state 
of the $J_1$-$J_2$ model only for $0.35\, J_1  \lesssim J_2 \lesssim 0.50\, J_1$, see Sec.~\ref{sec:introduction}.
As pointed out in Ref.~\cite{buzo24}, such a feature of the harmonic approximation was observed by one of us in previous studies \cite{doretto14,doretto20,leite19}.
Moreover, the $D^{\rm max}_\parallel$ below which the columnar VBS phase is stable 
assumes the largest value for $J_2 \approx 0.20\, J_1$ and monotonically decreases  with $J_2$. Note that $J_2 \approx 0.20\, J_1$ is roughly the (lower) border of the paramagnetic region according 
to the DMRG and Schwinger boson mean-field theory results.

Figures~\ref{fig:params1}(a) and \ref{fig:params1}(b), respectively, show the 
behaviour of the parameters $N_0$ and $\mu$ as functions of the DM interaction $D_\parallel$,
for fixed values of next-nearest-neighbor 
exchange coupling $J_2$ and the external magnetic field $h_z = -0.05\, J_1$. 
Note that both data including [Eq.~\eqref{eq:coef-C2} with $\eta = 1$] and neglecting [Eq.~\eqref{eq:coef-C2} with $\eta = 0$] 
the effects of the linear term in Eq.~\eqref{eq:ham-loc} are shown.
For fixed $J_2$, $N_0$ decreases monotonically as $D_\parallel$
increases, indicating that the singlet-condensate weight is reduced.  This
trend agrees with the square-lattice results of Ref.~\cite{buzo24}.
As expected from Eq.~\eqref{eq:coef-cprime}, the corrections associated with the linear term \eqref{eq:ham-loc} 
increase with the DM interaction $D_\parallel$ and, in particular, are stronger for $J_2 < 0.35\, J_1$.

The ground-state energy \eqref{eq:EGS} is shown in Fig.~\ref{fig:params1}(c). 
Similar to the parameter $N_0$, it decreases as the DM interaction $D_{\parallel}$ 
increases for a given value of  $J_2$. It can also be seen that the corrections due to 
the linear term \eqref{eq:ham-loc} are more pronounced for $D_\parallel > 0.20\, J_1$, as also found 
for the behavior of the parameter $N_0$, see Fig.~\ref{fig:params1}(a).
For a fixed DM interaction $D_\parallel$ [Fig.~\ref{fig:params1}(d)], we find that $E_{GS}$ increases with the exchange coupling $J_2$ up to 
$J_2 \sim 0.30\, J_1$ and then it decreases, regardless of the value of the DM interaction.
The result for $D_\parallel = 0$ agrees well with a previous one for the $J_1$-$J_2$ model 
based on the bond-operator formalism (see dashed green line in Fig.~4 from Ref.~\cite{mulder10})
and is also quite close to the ground state energy calculated within a Schwinger boson mean-field theory 
(see Fig.~6 from Ref.~\cite{zhang13}).

A few remarks concerning the procedure employed to treat the linear term \eqref{eq:ham-loc} are in order here. As discussed in Sec.~\ref{subsec:dzyaloshinskii-moriya-effective-hamiltonian}, we included corrections due to the presence of the linear term up to first order in the parameter $D_\parallel/(2J_1)$. For the columnar VBS phase on the honeycomb lattice, such a procedure should be employed only for $D_\parallel \lesssim 0.20\, J_1$, since, as exemplified in Figs.~\ref{fig:params1}(a) and \ref{fig:params1}(c), only in this region are the corrections indeed small. 
Therefore, we show the region of stability of the columnar VBS (Fig.~\ref{fig:phase-diagram}) calculated without the effects of the linear term [Eq.~\eqref{eq:coef-C2} with $\eta = 0$] and, in the following, we will focus on the parameter region $D_\parallel < 0.20\, J_1$ and discuss properties of the columnar VBS including the effects associated with the linear term [Eq.~\eqref{eq:coef-C2} with $\eta = 1$].

\begin{figure*}[t]
    \centering
    \begin{subfigure}[b]{0.40\textwidth}
        \centering
        \llap{\textbf{(a)}\hspace{-0.05em}}
        \vspace*{-1.5em}
        \includegraphics[width=\textwidth]{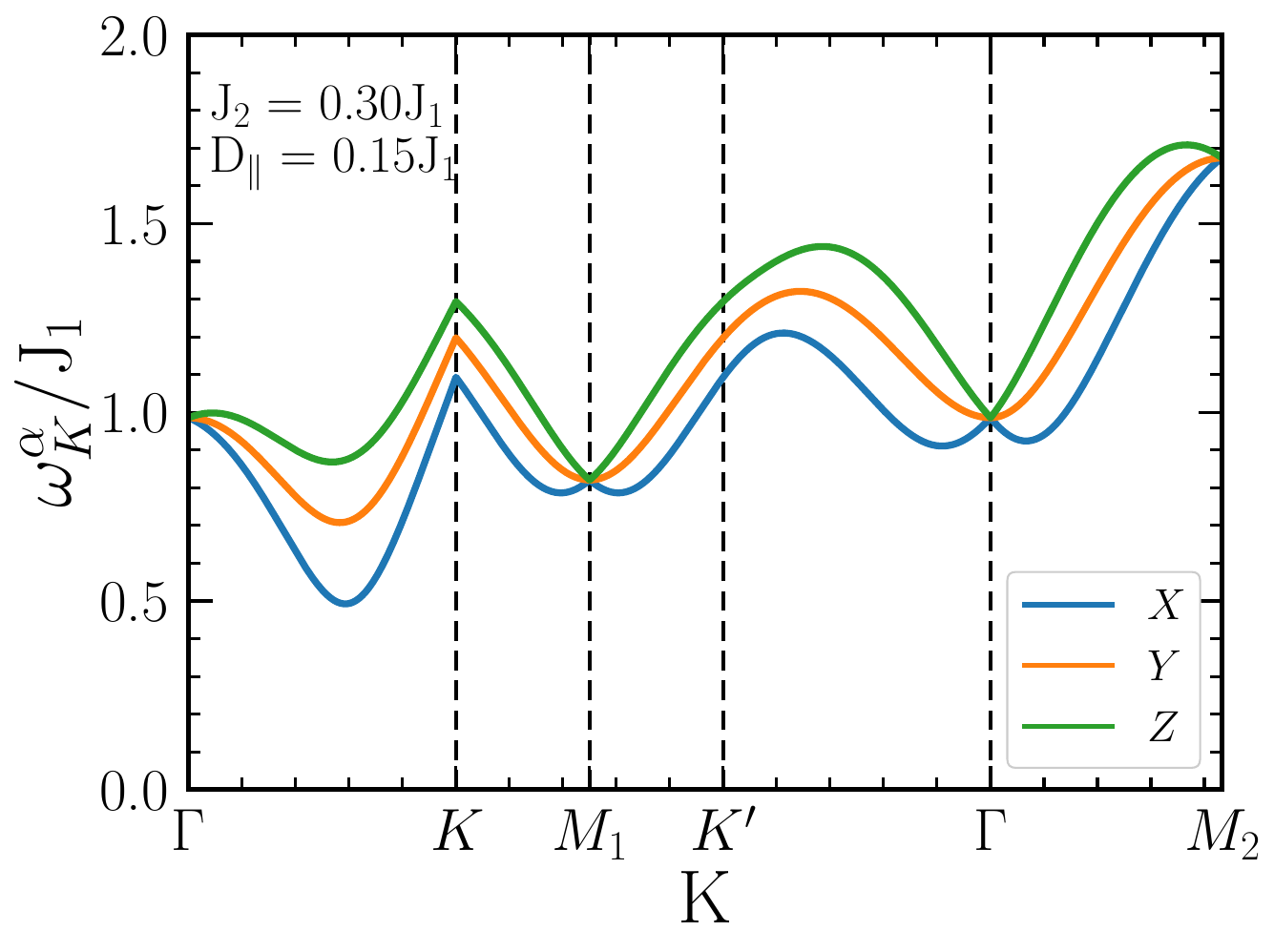}
    \end{subfigure}
    \hspace{1.5cm}
    \begin{subfigure}[b]{0.40\textwidth}
        \centering
        \llap{\textbf{(b)}\hspace{-0.05em}}
        \vspace*{-1.5em}
        \includegraphics[width=\textwidth]{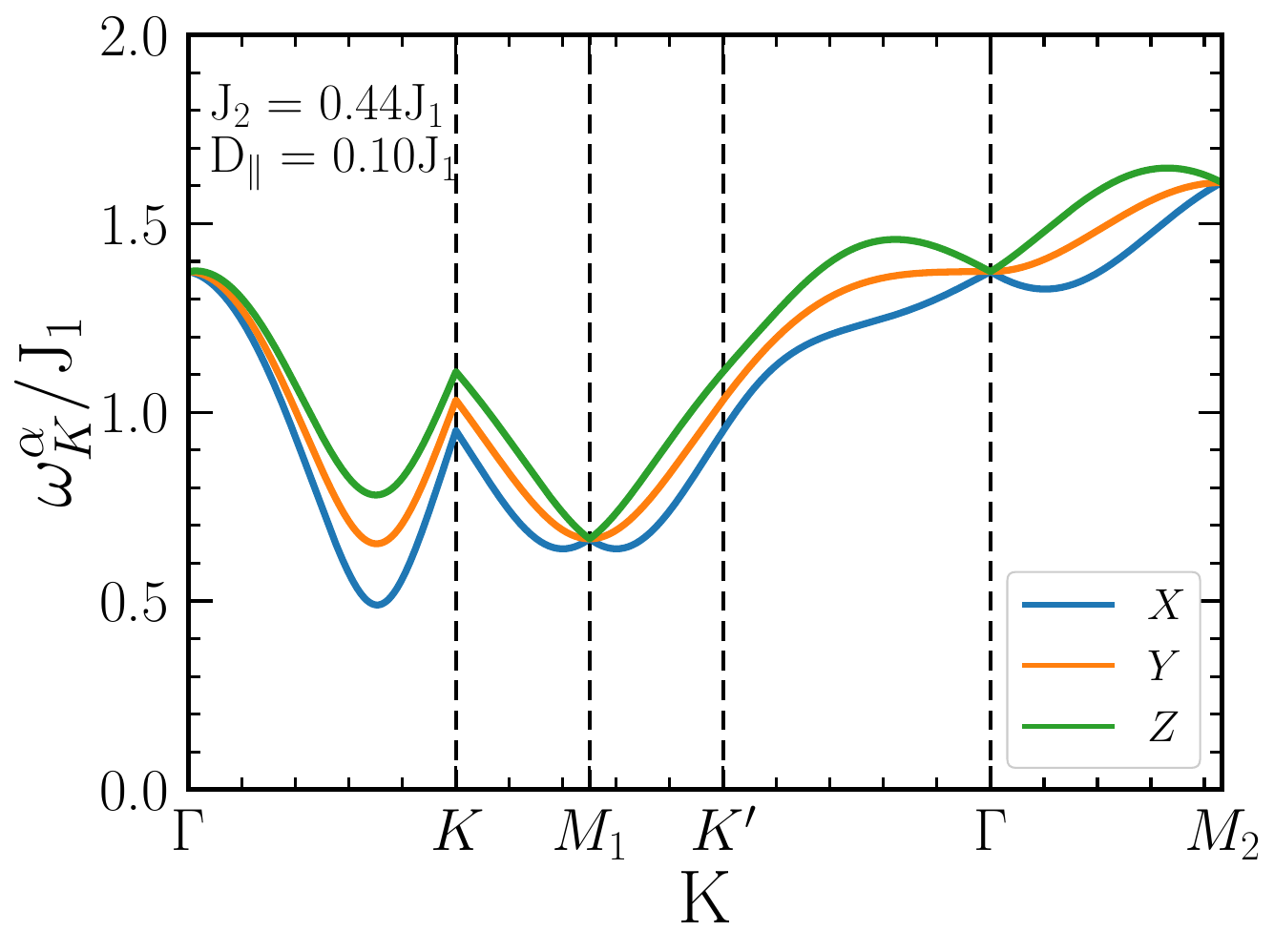}
    \end{subfigure}

    \bigskip
    \begin{subfigure}[b]{0.40\textwidth}
        \centering
        \llap{\textbf{(c)}\hspace{-0.05em}}
        \vspace*{-1.5em}
        \includegraphics[width=\textwidth]{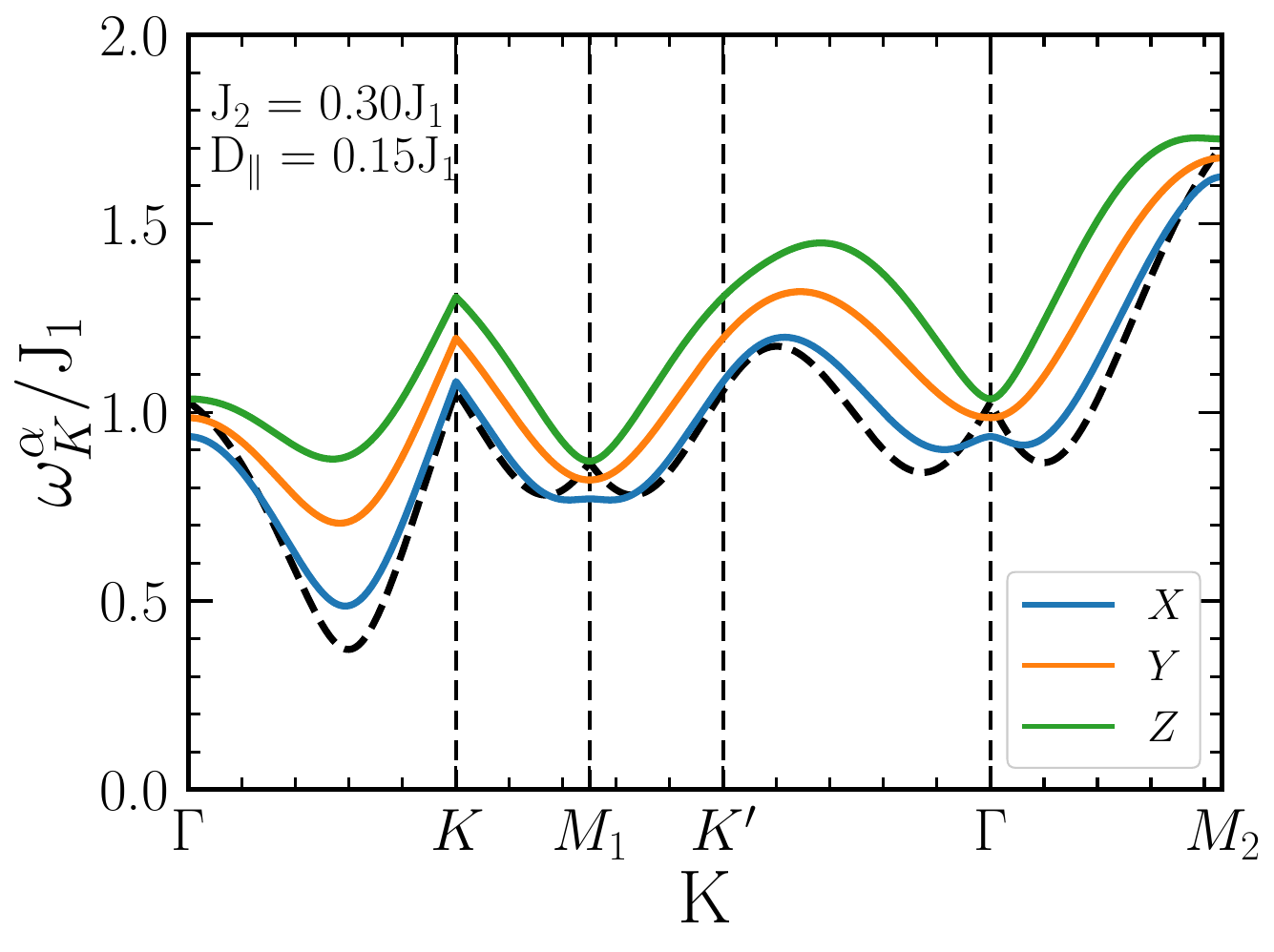}
    \end{subfigure}
    \hspace{1.5cm}
    \begin{subfigure}[b]{0.40\textwidth}
        \centering
        \llap{\textbf{(d)}\hspace{-0.05em}}
        \vspace*{-1.5em}
        \includegraphics[width=\textwidth]{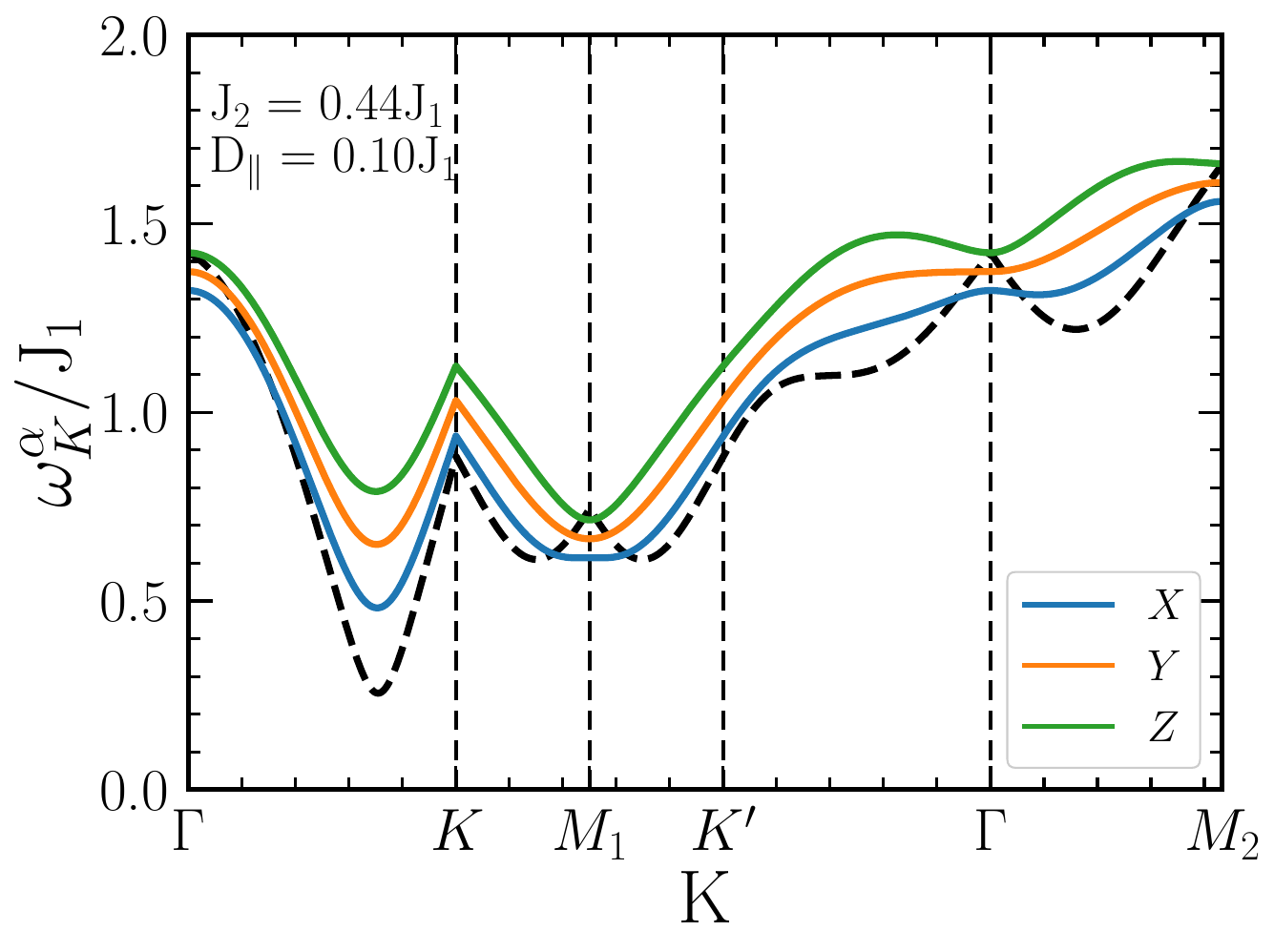}
    \end{subfigure}
    \caption{\justifying{Triplon dispersion relations 
    $\wk^x$ (solid blue line),  $\wk^y$ (solid orange line), and  $\wk^z$ (solid green line) 
    of the columnar VBS phase of Heisenberg model \eqref{eq:ham} within the harmonic approximation, along a path in the dimerized Brillouin zone [Fig.~\ref{fig:model}(d)]:
    (a) and (c) $J_2 = 0.30 \, J_1$ and $D_\parallel = 0.15 \, J_1$; 
    (b) and (d) $J_2 = 0.44 \, J_1$ and $D_\parallel = 0.10 \, J_1$. 
    In the upper panels, the external magnetic field is set to zero, while in the lower ones, it is fixed at $h_z = -0.05 \, J_1$. 
    In the lower panels, the dashed black line is the (degenerate) triplon dispersion relation for the Heisenberg model \eqref{eq:ham} in the absence of both DM interaction and external magnetic field ($D_\parallel = h = 0$).}}
    \label{fig:dispersion}
\end{figure*}

\begin{figure*}[t]
    \centering
    \begin{subfigure}[b]{0.45\textwidth}
        \centering
        \llap{\textbf{(a)}\hspace{-0.05em}}
        \vspace*{-1.5em}
        \includegraphics[width=\textwidth]{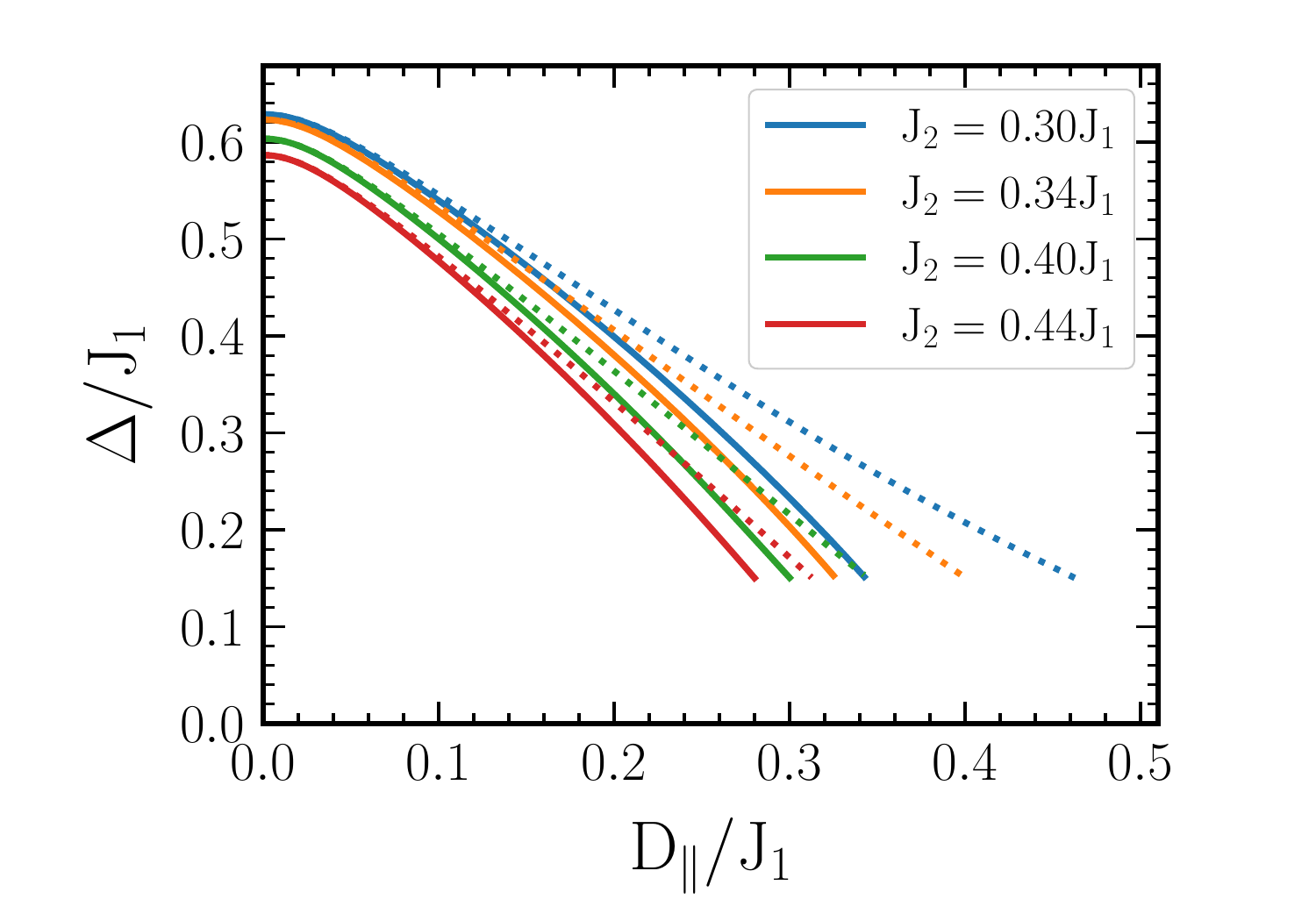}
    \end{subfigure}
    \hspace{0.7cm}
    \begin{subfigure}[b]{0.45\textwidth}
        \centering
        \llap{\textbf{(b)}\hspace{-0.05em}}
        \vspace*{-1.5em}
        \includegraphics[width=\textwidth]{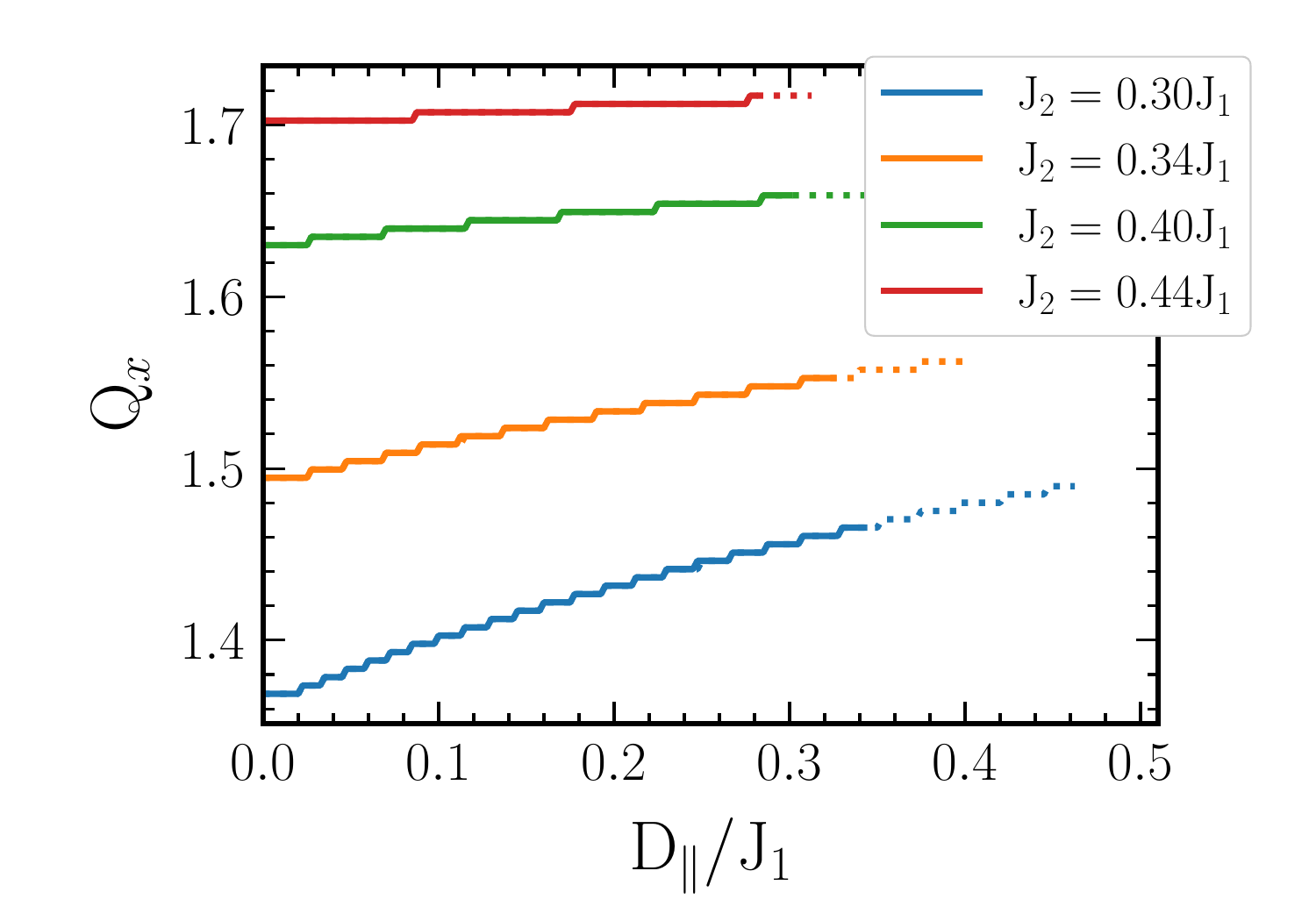}
    \end{subfigure}
    \caption{\justifying{
    The behaviour of 
    (a) the triplon excitation gap $\Delta / J_1$ and 
    (b) the $x$-component $Q_x$ of the momentum $\mathbf{Q} = (Q_x,0)$ associated with the triplon gap
    on the DM interaction $D_\parallel/J_1$ for a fixed external magnetic field $h_z = -0.05 \, J_1$.
    Next-nearest-neighbor couplings $J_2^B = 0.9 J_2^A \equiv 0.9 J_2$ [Eq.~\eqref{eq:j2b}], with
    $J_2 = 0.30 \, J_1$ (blue line), $J_2 = 0.34 \, J_1$ (orange line), 
    $J_2 = 0.40 \, J_1$ (green line), and $J_2 = 0.44 \, J_1$ (red line);
    the solid and dotted lines indicate, respectively, results neglecting [Eq.~\eqref{eq:coef-C2} with $\eta = 0$]  
    and including [Eq.~\eqref{eq:coef-C2} with $\eta = 1$] the effects of the linear term \eqref{eq:ham-loc}.
   } }
    \label{fig:params2}
\end{figure*}

\subsection{Parallel DM vector pattern: Triplon dispersion relations}
\label{subsec:triplon-energies}

Figure~\ref{fig:dispersion} shows the triplon dispersion relations $\omega^\alpha_\bk$ 
for $J_2 = 0.30 \, J_1$ and $D_\parallel = 0.15 \, J_1$, with external magnetic field
$h_z = 0$ [Fig.~\ref{fig:dispersion}(a)] and $h_z = - 0.05 \, J_1$ [Fig.~\ref{fig:dispersion}(c)],
as well as 
for $J_2 = 0.44 \, J_1$ and $D_\parallel = 0.10 \, J_1$, with 
$h_z = 0$ [Fig.~\ref{fig:dispersion}(b)] and $h_z = - 0.05 \, J_1$ [Fig.~\ref{fig:dispersion}(d)].
Here the corrections due to the linear term \eqref{eq:ham-loc} are taken into account and,
in the lower panels, the dashed black lines are the (degenerate) 
triplon energies in the absence of the DM interaction and external magnetic field
($D = h = 0$), included here for comparison.
We see that the triplon excitation spectra are gapped, a feature of a VBS phase.
Moreover, we note that the degeneracy of the three triplon bands, present in the 
excitation spectrum of the $J_1$-$J_2$ model (dashed black lines in the lower panels), 
is only partially lifted by the DM interaction, as the triplon bands still touch at the 
$\mathbf{\Gamma}$ and $\mathbf{M}_i$ points of the (dimerized)  
Brillouin zone, see Fig.~\ref{fig:model}(d).  
Although the DM interaction is essential for inducing phases with topologically nontrivial 
excitation bands \cite{review-zhang}, here, it is insufficient, on its own, 
to lead to well separated bands in which their Chern numbers can be properly defined. 
Therefore, an external magnetic field is introduced to fully lift the degeneracy between the
three triplon bands, as illustrated in Figs.~\ref{fig:dispersion}(c) and \ref{fig:dispersion}(d).
It should be  noted that a finite magnetic field was included in studies of magnons in magnets with
long-range order \cite{mook14,owerre16,malki19,hotta19} and triplons 
on the square \cite{buzo24}, Shastry-Sutherland \cite{penc15, malki17}, and 
maple-leaf and star \cite{esaki25} lattices.

Figure~\ref{fig:dispersion} also shows that the triplon excitation gap $\Delta$, i.e., the minimum of the 
$\omega^x_\bk$ triplon band, is associated with an incommensurate momentum $\bQ$: 
For $J_2 = 0.30 \, J_1$ and $D_\parallel = 0.15 \, J_1$, we find that $\bQ = (1.4172,0)$,
while,
for $J_2 = 0.44 \, J_1$ and $D_\parallel = 0.10 \, J_1$, $\bQ = (1.7074,0)$,
regardless of the value of the  magnetic field $h_z$.
The incommensurate value of $\bQ$ here seems to be a remnant of the spiral
phase that sets in for $J_2/J_1 > 1/6$ in the classical phase diagram (spin $S \rightarrow \infty$)
of the $J_1$-$J_2$ model \cite{fouet01,baez17,hikaru19,consoli24}. As mentioned in Sec.~\ref{sec:introduction}, 
for $1/6 < J_2/J_1 < 1/2$, the spiral ordering wave vector $\bQ_0$ spans a closed contour around the center of the
Brillouin zone; for $J_2 = 0.30\, J_1$ and $J_2 = 0.44 \, J_1$, the spiral ordering wave
vectors along the $\Gamma$-$K$ line are respectively given by 
$\bQ_0 = (1.4214,0)$ and $\bQ_0 = (1.7350,0)$.

The triplon gap $\Delta$ as a function of the DM interaction $D_\parallel$ for a given exchange coupling $J_2$ 
is shown in Fig.~\ref{fig:params2}(a). The behavior of $\Delta$ is qualitatively the same 
for all values of the coupling $J_2$. In each case, the gap decreases as $D_\parallel$ increases. 
In the absence of the corrections due to the linear term \eqref{eq:ham-loc}  (solid lines), the smallest value of the
triplon gap $\de{\Delta_{\textrm{min}} \sim 0.15}$ is reached at the DM value $D^{\rm max}_\parallel$ above which a
numerical solution for the self-consistent equations \eqref{eq:self-const} can no longer be found, see solid black
line in Fig.~\ref{fig:phase-diagram}. 
Such a feature was previously observed by one of us in a study of a square lattice AFM,
see Fig.~2(c) from Ref.~\cite{buzo24}.
Including the corrections due to the linear term (dashed lines), we find that such corrections get more pronounced for $D_\parallel \gtrsim 0.20\, J_1$, as observed for the parameter $N_0$ and the ground-state energy in Figs.~\ref{fig:params1}(a) and \ref{fig:params1}(c), respectively.
Finally, concerning the behavior of the momentum $\bQ$ associated with the triplon gap $\Delta$ [Fig.~\ref{fig:params2}(b)],
we find that the effect of the DM interaction seems to be rather small.

The behavior of the triplon gap $\Delta$ and the momentum $\bQ$ seem to indicate that a
quantum phase transition from a columnar VBS phase to a spiral one with long-range magnetic order
takes place at $D_\parallel \sim D^{\rm max}_\parallel$. Whether it is a first-order quantum phase
transition or a continuous one is a separate issue that lies outside the scope of the present work. 
Again, such features were also found in a previous study concerning the square-lattice $J_1$-$J_2$ 
model \cite{buzo24}.


\section{Berry curvatures and Chern numbers}
\label{sec:berry-curvatures-and-chern-numbers}

\begin{figure*}[t]
    \centering
    \begin{subfigure}[b]{1\textwidth}
        \centering
        \caption{$J_2 = 0.30 \, J_1$, $D_\parallel = 0.15 \, J_1$, and $h_z = - 0.05 \, J_1$}
        \includegraphics[width=\textwidth]{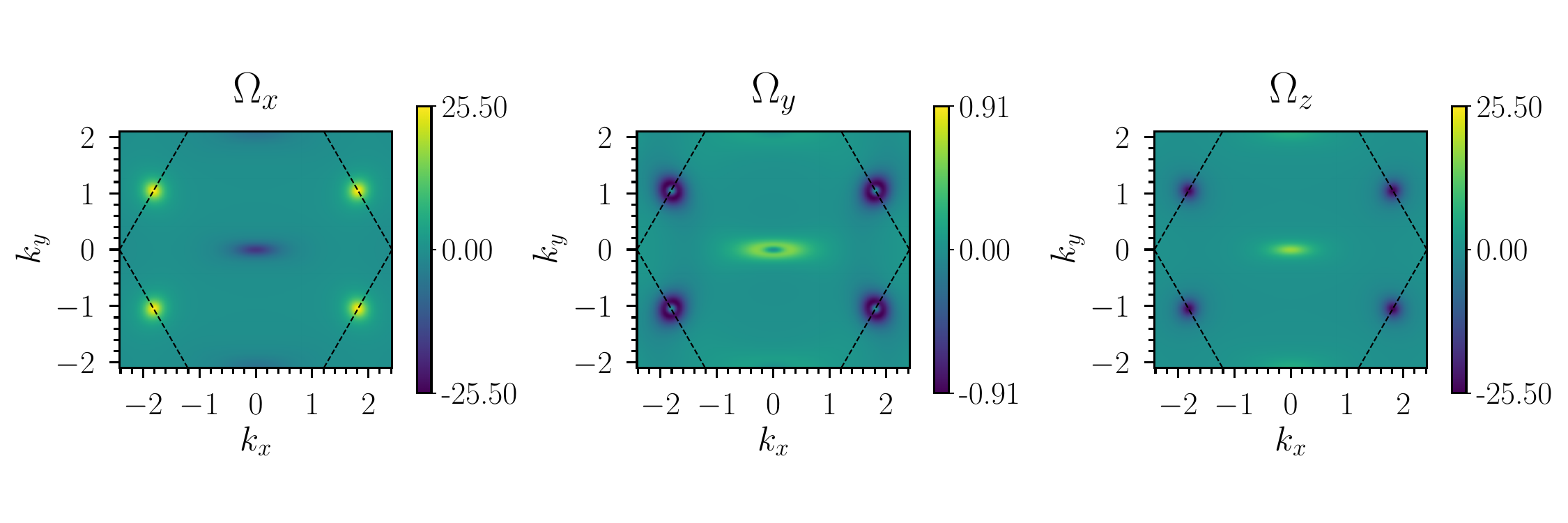}
    \end{subfigure}
    %
    \vskip0.5cm    
    \begin{subfigure}[b]{\textwidth}
        \centering
        \caption{$J_2 = 0.44 \, J_1$, $D_\parallel = 0.10 \, J_1$, $h_z = - 0.05 \, J_1$}
        \includegraphics[width=\textwidth]{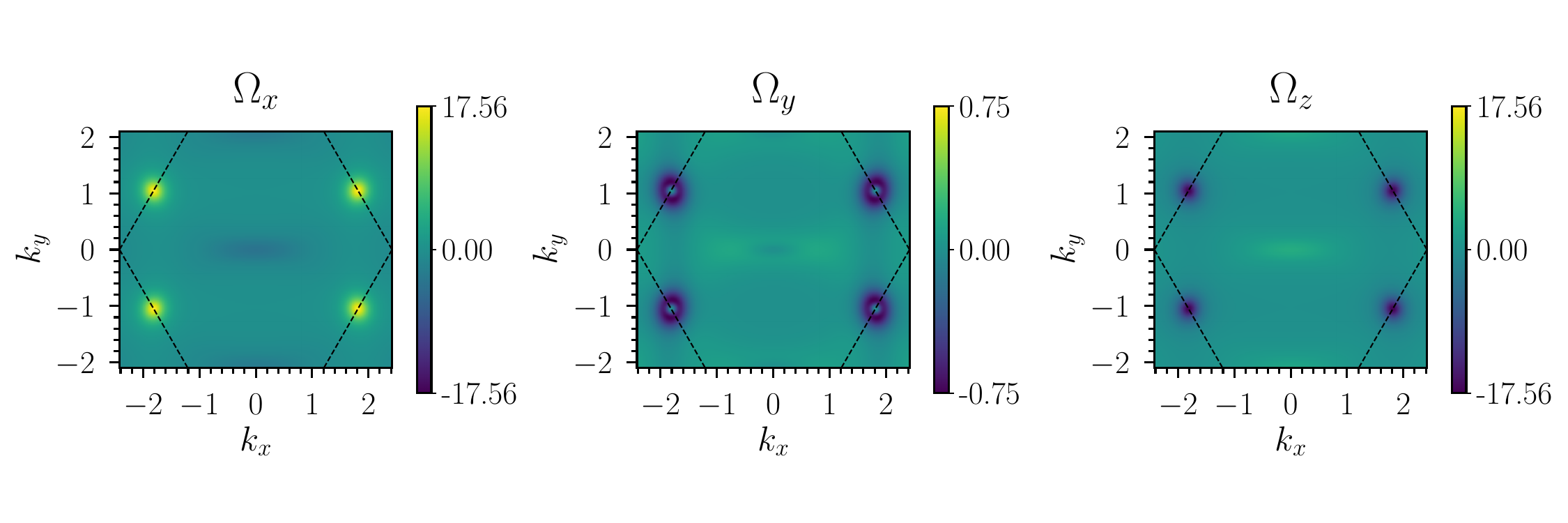}
    \end{subfigure}
    \caption{\justifying{Berry curvatures $\Omega_{\alpha} (\bk)$ [see Eq.~\eqref{eq:berry-curvature}] of the triplon excitation bands $\wk^{\alpha}$ displayed in Figs.~\ref{fig:dispersion}(c) and \ref{fig:dispersion}(d). The next-nearest-neighbor exchange and DM couplings are 
    (a) $J_2 = 0.30 \, J_1$ and $D_\parallel = 0.15 \, J_1$; 
    (b) $J_2 = 0.44 \, J_1$ and $D_\parallel = 0.10 \, J_1$. 
    The external magnetic field is fixed at $h_z = -0.05 \, J_1$.}}
    \label{fig:berry-curvature}
\end{figure*}

In this section, we explore the topological aspects of the columnar VBS on a honeycomb lattice, in particular, the properties 
of the triplon excitations. The idea is to determine whether 
the nearest-neighbor DM interaction \eqref{eq:ham-dm}, with an in-plane parallel vector pattern $D_\parallel$, 
induces a nontrivial topology on the triplon excitation bands of the columnar VBS phase of the model \eqref{eq:ham}.

\begin{figure*}[t]
    \centering
    \begin{subfigure}[b]{0.45\textwidth}
        \centering
        \llap{\textbf{(a)}\hspace{-0.05em}}
        \vspace*{-1.5em}
        \includegraphics[width=\textwidth]{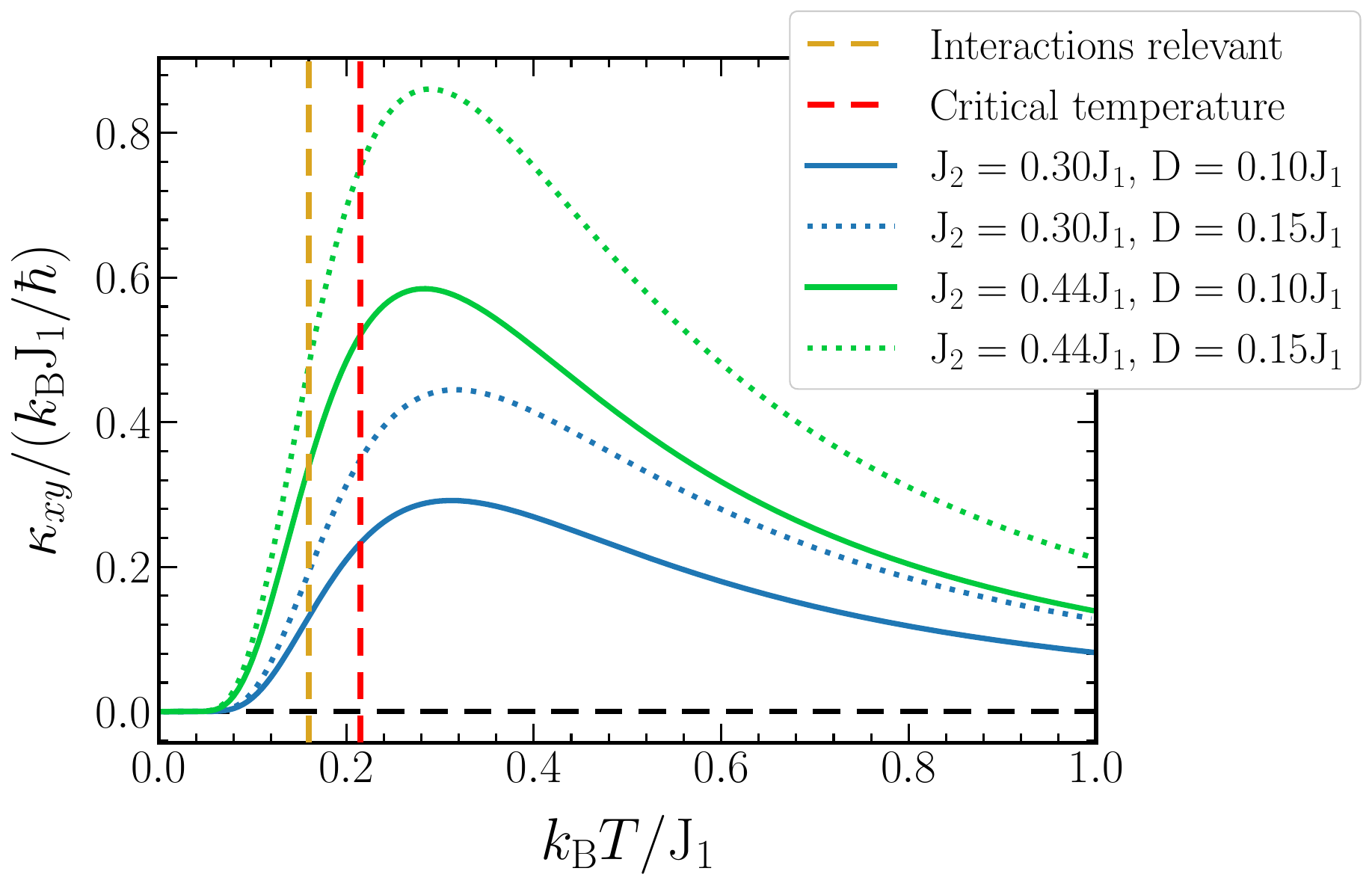}
    \end{subfigure}
    \hspace{0.75cm}
    \begin{subfigure}[b]{0.435\textwidth}
        \centering
        \llap{\textbf{(b)}\hspace{-0.05em}}
        \vspace*{-1.5em}
            \includegraphics[width=\textwidth]{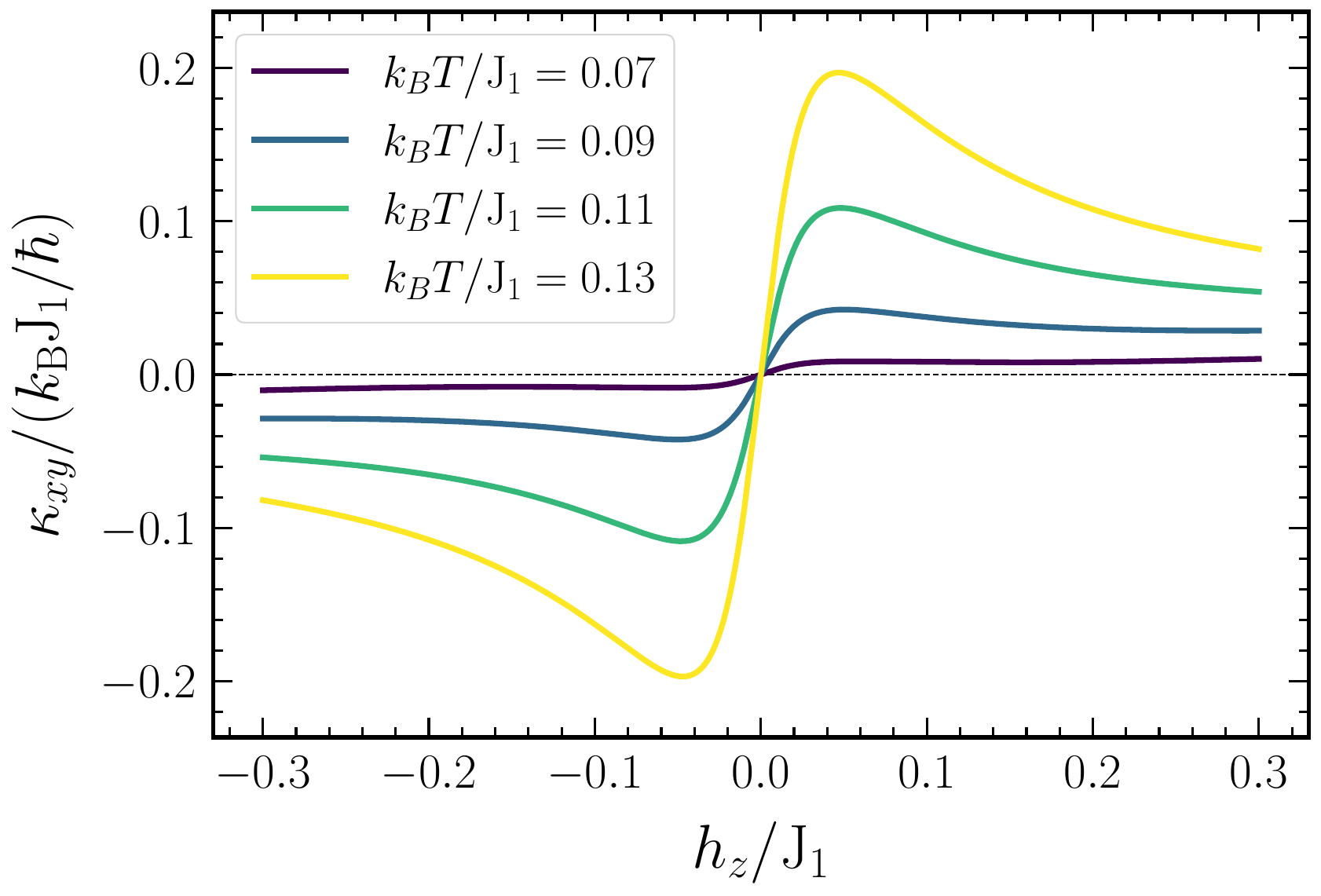}
    \end{subfigure}

    \caption{\justifying{
    (a) Thermal Hall conductivity $\kappa_{xy} / \de{k_B J_1 / \hbar}$ as a function of temperature $k_B T/ J_1$ for the Heisenberg model \eqref{eq:ham} associated with the triplon excitations of the columnar VBS phase. 
    Blue and green lines correspond to $J_2 = 0.30 \, J_1$ and $J_2 = 0.44 \, J_1$, respectively, 
    with the solid (dashed) lines indicating $D_\parallel = 0.10 \, J_1$ ($ D_\parallel = 0.15 \, J_1$). 
    The external magnetic field is fixed at $h_z = 0.05 \, J_1$. 
    The vertical dashed yellow line indicates the onset of relevant triplon-triplon interactions, while 
    the vertical dashed red line marks the critical temperature where the number of triplon excitations exceeds 
    the number of sites $N_{\mathcal{D}} = N /2$ of the dimerized lattice $\mD$ (see text for details). 
    (b) Thermal Hall conductivity as a function of the external magnetic field $h_z / \, J_1$ 
    for $J_2 = 0.44 \, J_1$ and $D_\parallel = 0.10 \, J_1$; 
    the temperatures are within the range $k_B T = 0.07$--$0.13 \, J_1$.}}
    \label{fig:thermal-hall}
\end{figure*}

For a two-dimensional system, the Berry curvature $\Omega_\alpha(\bk)$ of an excitation band $\omega^\alpha_\bk$ determined from a  quadratic bosonic Hamiltonian with anomalous terms, such as the Hamiltonian \eqref{eq:ham-matrix},
is given by \cite{prl-ryo11, prb-ryo11, ryo13, ryo14}
\begin{equation}
    \label{eq:berry-curvature}
    \Omega_\alpha \de{\bk} = i \epsilon_{ \mu \nu } \De{ \Ib \pp{\Tkd}{k_{\mu}} \Ib \pp{\Tk}{k_{\nu}} }_{\alpha\alpha}.
\end{equation}
Here, $\alpha = 1,2,...,6$, $\mu, \nu = x, y$,
$\epsilon_{ \mu \nu }$ is the completely antisymmetric tensor with $\epsilon_{xy} = 1$,
the $6 \times 6$ matrices $\Ib$ and $\Tk$ are given by Eqs.~\eqref{eq:Ib} and \eqref{eq:Tmatrix}, respectively, 
and $\De{\;}_{\alpha\alpha}$ denotes the $\alpha$-th diagonal element of the corresponding square matrix. 
The form of the diagonal Hamiltonian~\eqref{eq:diag-ham} implies that the Berry curvatures associated with the triplon bands 
$\wk^x$, $\wk^y$, and $\wk^z$ are as follows: 
$\Omega_x \de{\bk} \equiv \Omega_1 \de{\bk} = \Omega_4 \de{\bk}$, 
$\Omega_y \de{\bk} \equiv \Omega_2 \de{\bk} = \Omega_5 \de{\bk}$, 
and $\Omega_z \de{\bk} \equiv \Omega_3 \de{\bk} = \Omega_6 \de{\bk}$.
The Berry curvatures \eqref{eq:berry-curvature} were calculated numerically  from the transformation matrices
\eqref{eq:Tmatrix} determined according to the procedure outlined in Sec.~\ref{subsec:diagonalization-of-the-quadratic-hamiltonian}.

The Berry curvatures of the triplon  bands $\wk^\alpha$ for   
$J_2 = 0.30 \, J_1$, $D_\parallel = 0.15 \, J_1$, and $h_z = - 0.05 \, J_1$ [see Fig.~\ref{fig:dispersion}(c)]
are displayed in Fig.~\ref{fig:berry-curvature}(a), 
while the ones for
$J_2 = 0.44 \, J_1$ and $D_\parallel = 0.10 \, J_1$, and $h_z = - 0.05 \, J_1$ [see Fig.~\ref{fig:dispersion}(d)]
are shown in Fig.~\ref{fig:berry-curvature}(b). 
We see that the Berry curvatures are zero throughout most of the dimerized Brillouin zone [Fig.~\ref{fig:model}(d)],
but they acquire notably finite values in the regions surrounding the  $\mathbf{\Gamma}$, $\mathbf{M_1}$ and $\mathbf{M_3}$ points; the peaks of $\Omega_x(\bk)$ and $\Omega_z(\bk)$ are significantly more intense than those of $\Omega_y(\bk)$ and the (approximate) antisymmetric relation $\Omega_{x} (\bk) \simeq -\Omega_{z} (\bk)$ holds. 
Furthermore, we find that the peak intensities decrease as $D_{\parallel} \rightarrow 0$, while their positions remain unaffected.
This feature indicates that the DM interaction with a parallel vector pattern gives rise to a finite Berry curvature for the triplon bands.
Our findings for the Berry curvature described above are qualitatively similar to the ones determined for the columnar VBS phase on a square-lattice AFM \cite{buzo24}. In that case, a DM vector pattern parallel to the lattice plane was also considered in addition to an external magnetic field $\bB = B\hat{z}$.

Finally, we calculate the Chern number $C_\alpha$ for each triplon band $\wk^\alpha$.
This quantity is defined as the integral of the Berry curvature $\Omega_\alpha(\bk)$ 
over the dimerized Brillouin zone \cite{ryo13,ryo14},
\begin{equation}
    \label{eq:chern-number}
    C_{\alpha} = \frac{1}{2 \pi} \int_{\textrm{BZ}} d^2k \; \Omega_{\alpha} \de{\bk}.
\end{equation}
Considering the Berry curvatures displayed in Figs.~\ref{fig:berry-curvature}(a) and 
 \ref{fig:berry-curvature}(b) and a numerical evaluation 
 of the integral in Eq.~\eqref{eq:chern-number}, 
it is found that, apart from small deviations from zero due to finite numerical precision, the Chern numbers of the three triplon bands vanish for both parameter 
sets of the model \eqref{eq:ham}.
As shown in Fig.~\ref{fig:berry-curvature}, this outcome arises from a cancellation 
between sharp, high-amplitude peaks of one sign and broad, low-amplitude ones with the opposite sign.
We investigated different parameter sets for the model \eqref{eq:ham} and found that this is indeed a
feature of the columnar VBS phase on the honeycomb lattice. 
Therefore, similarly to the columnar VBS phase on a square-lattice AFM \cite{buzo24}, it is found that the 
triplon excitation bands $\wk^\alpha$ of the columnar VBS phase on the honeycomb-lattice AFM \eqref{eq:ham} 
are topologically trivial, despite the fact that the DM interaction does give them a finite Berry curvature.

\section{Thermal Hall conductivity}
\label{sec:thermal-hall-conductivity}

We finally consider the transport properties of the columnar VBS phase of the honeycomb-lattice AFM \eqref{eq:ham}, focusing on the determination of the thermal Hall conductivity $\kappa_{xy}$ due to the triplon excitations. 
We follow a well-established framework \cite{prl-ryo11,prb-ryo11} that was derived via both semi-classical and linear-response theories and was initially employed to calculate $\kappa_{xy}$ due to magnon excitations in long-range ordered 
FMs \cite{ryo13, ryo14}.
The thermal Hall conductivity $\kappa_{xy}$ due to triplons on 
Shastry-Sutherland lattice AFMs \cite{penc15,malki17,coldea17} and on a square-lattice AFM \cite{buzo24} were also determined within this framework.  

For a system described by non-interacting bosonic excitations, the thermal Hall conductivity $\kappa_{xy}$ assumes the form 
\cite{prl-ryo11,prb-ryo11} 
\begin{equation}
    \label{eq:thermal-hall}
    \kappa_{xy} = -\frac{k^2_{B} T}{\hbar V} \sum_{\bk, \alpha} c_2 \De{f_B \de{ \wk^{\alpha} } } \Omega_{\alpha}(\bk).
\end{equation}
Here, $k_B$ is the Boltzmann constant, 
$T$ is the temperature, $V$ is the sample area, $f_B \de{\epsilon}$ is the Bose-Einstein distribution function,
\begin{equation}
\label{eq:bose-einstein}
f_B \de{\epsilon} = \frac{1}{e^{\epsilon/k_B T} - 1},
\end{equation}
and $\Omega_{\alpha} (\bk)$ is the Berry curvature \eqref{eq:berry-curvature} corresponding to the triplon band 
$\wk^{\alpha}$ (with $\alpha = x, y, z$). The function $c_2 (x)$ is given by
\begin{align}
    c_2 \de{x} &= \int_{0}^{x} dt \, \de{ \ln{\frac{1 + t}{t}} }^2
    \nonumber \\
    &= \de{1 + x} \de{ \ln{ \frac{1 + x}{x} } }^2 - \de{\ln{x}}^2 - 2 \textrm{Li}_2 \de{-x},
    \label{eq:c2}
\end{align} 
where $\textrm{Li}_2(x)$ is the polylogarithm  defined as
\begin{equation}
    \label{eq:polilog}
    \textrm{Li}_n \de{z} = \sum_{j = 1}^{\infty} z^j/j^n.
\end{equation}
We used the Berry curvatures determined in the previous section and calculated numerically  the thermal Hall conductivity \eqref{eq:thermal-hall}.

Figure~\ref{fig:thermal-hall}(a) shows the temperature dependence of the thermal Hall conductivity $\kappa_{xy}$
for model parameters $J_2 = 0.30$ and $0.44\, J_1$, 
DM interaction $D_\parallel = 0.10$ and $0.15 \, J_1$, and 
external magnetic field of $h_z = 0.05 \, J_1$. 
Regardless the values of the model parameters, one finds that $\kappa_{xy}$ has a peak around $T \sim 0.35 \, J_1 / k_B$, which increases with the DM interaction $D_\parallel$ for a given coupling $J_2$. In addition, $\kappa_{xy}$ vanishes in the low-temperature region, a feature directly related to the finite triplon gap in the excitation spectrum, see Fig.~\ref{fig:dispersion}.
One should also mention that the temperature dependence of $\kappa_{xy}$ found here is in (qualitative) agreement with the general behavior for the thermal Hall conductivity phenomenologically derived in Ref.~\cite{yang20}.

Figure~\ref{fig:thermal-hall}(b) illustrates the dependence of $\kappa_{xy}$ on the external magnetic field $h_z$ for several temperatures, with fixed exchange coupling $J_2 = 0.44 \, J_1$ and DM interaction $D_\parallel = 0.10 \, J_1$. 
The thermal Hall conductivity exhibits two key features related to the external field. First, it is an odd function of the magnetic field, satisfying $\kappa_{xy}(-h_z) = -\kappa_{xy}(h_z)$, which implies that the direction of the transverse heat current changes as the direction of the magnetic field is reversed.
Besides, $\kappa_{xy}$ is a non-monotonic function of the magnetic field, since (for $h_z > 0$) it increases from zero up to 
$h_z \sim 0.05 \, J_1$ and then it seems to decrease asymptotically towards zero for larger fields. 
Moreover, we see that the temperature dependence of $\kappa_{xy}$ in Fig.~\ref{fig:thermal-hall}(b) agrees with the one found in Fig.~\ref{fig:thermal-hall}(a):
As the temperature increases, so does the thermal Hall conductivity, for any value of $h_z$.
Finally, in the absence of the field, the thermal Hall conductivity is not observed, as expected~\cite{review-zhang}.

\begin{figure}[b]
    \centering
    \includegraphics[width=0.90\linewidth]{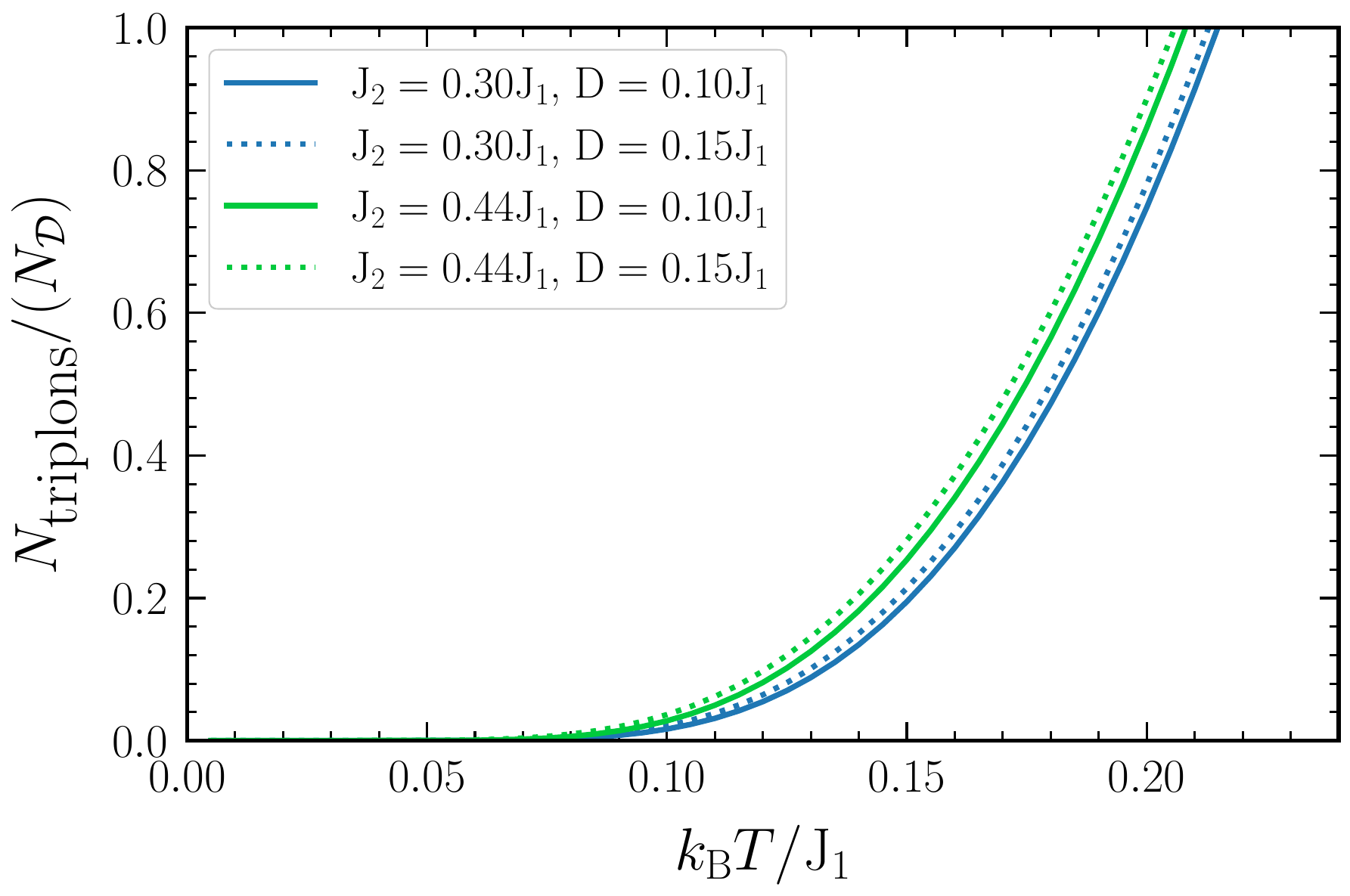}
    \caption{\justifying{Number of triplons $N_{\textrm{triplons}}/N_{\mathcal{D}}$ as a function of the temperature $k_B T / J_1$ for the columnar VBS phase of the model \eqref{eq:ham}: 
    $J_2 = 0.30 \, J_1$ and $D_\parallel = 0.10 \, J_1$ (solid blue line);
    $J_2 = 0.30 \, J_1$ and $D_\parallel = 0.15 \, J_1$ (dashed blue line);
    $J_2 = 0.44 \, J_1$ and $D_\parallel = 0.10 \, J_1$ (solid green line);
    $J_2 = 0.44 \, J_1$ and $D_\parallel = 0.15 \, J_1$ (dashed green line). 
    The external magnetic field is fixed at $h_z = 0.05 \, J_1$ and
    $N_{\mathcal{D}}$ is the number of sites of the dimerized lattice $\mD$. }}
    \label{fig:n-triplons}
\end{figure}

A few words concerning the temperature range where the approximations employed here are appropriate for the determination of the thermal Hall conductivity are in order here.
The total number of triplons, $N_{\textrm{triplons}}$, is given by
\begin{equation}
\label{eq:n-triplon}
N_{\textrm{triplons}} = \sum_{\bk, \alpha} \mean{ \bd_{\bk \alpha} b_{\bk \alpha} } 
                      = \sum_{\bk, \alpha} f_B \de{ \wk^{\alpha} },
\end{equation}
where $f_B(\epsilon)$ is the Bose-Einstein distribution function \eqref{eq:bose-einstein}
and $\wk^{\alpha}$ is the triplon energy \eqref{eq:Hkdiag}.
The temperature dependence of $N_{\textrm{triplons}}$ is plotted in Fig.~\ref{fig:n-triplons} for exchange couplings 
$J_2 = 0.30$ and $0.44 \, J_1$, DM interactions $D_\parallel = 0.10$ and $0.15 \, J_1$, and external magnetic field 
$h_z = 0.05 \, J_1$.
Based on the behavior of $N_{\textrm{triplons}}$,
we estimate that the triplon-triplon interactions arising from the cubic [Eqs.~\eqref{eq:fourier-ham3-j1j2} and \eqref{eq:fourier-ham3-DM}] and quartic [Eqs.~\eqref{eq:fourier-ham4-j1j2} and \eqref{eq:fourier-ham4-DM}] terms 
might become relevant for $T \gtrsim 0.15 \, J_1 / k_B$ 
[golden vertical dashed line in Fig.~\ref{fig:thermal-hall}(a)],
since the number of triplons exceeds $N_{\textrm{triplons}} \sim 0.25 \, N_\mathcal{D}$. 
Recall that triplon-triplon interactions are neglected within the harmonic approximation.
Moreover, for temperatures $T \sim 0.22 \, J_1 / k_B$ [red vertical dashed line in Fig.~\ref{fig:thermal-hall}(a)], the number of triplons $N_{\textrm{triplons}}$ approaches the number of sites $N_\mathcal{D} = N/2$ of the dimerized lattice $\mD$, indicating that the columnar VBS phase might  no longer be stable.

In summary, although the integral of the Berry curvature itself vanishes for each triplon band ($C_{\alpha}=0$), the thermal Hall conductivity \eqref{eq:thermal-hall} for the columnar VBS phase of the honeycomb-lattice AFM \eqref{eq:ham} is finite. This result follows from the fact that the integrand in Eq.~\eqref{eq:thermal-hall} is multiplied by the thermal weighting function $c_2 \De{f_B \de{\wk^{\alpha}}}$ and, since the triplon energy $\wk^{\alpha}$ varies across the Brillouin zone, this weighting is non-uniform, giving different statistical weights to the positive and negative regions of $\Omega_{\alpha} (\bk)$. As a result, the total sum does not vanish and yields a non-zero $\kappa_{xy}$.

\section{Summary and discussion}
\label{sec:summary-and-conlusions}

%
As in the square-lattice columnar VBS state studied in
Ref.~\cite{buzo24}, we found that an in-plane DM interaction generates finite Berry curvature
for the honeycomb-lattice triplon bands.  An out-of-plane magnetic field fully
lifts the band degeneracies, allowing a Chern number to be assigned to each
isolated band.  The numerically integrated Chern numbers vanish, whereas the
nonuniform thermal weighting of the local Berry curvature yields a finite
thermal Hall conductivity at nonzero low temperatures.
Such set of results for a columnar VBS in a honeycomb lattice seems to corroborate the fact that the {\sl no-go} condition \cite{lee10} for the thermal Hall effect does not apply to quantum paramagnetic phases: 
In addition to the square lattice VBS studied in Ref.~\cite{buzo24}, a thermal Hall effect due to spinons in a square lattice spin-liquid phase was also reported \cite{rhine19}.
We should recall (see Sec.~\ref{sec:introduction}) that the {\sl no-go} condition \cite{lee10} has been derived for a long-range ordered magnet within linear spin-wave theory and it indicates that a finite thermal Hall conductivity should not be observed in edge-sharing lattices, such as the square and honeycomb ones.
Interestingly, even for magnetic ordered phases, an exception was found: a finite thermal Hall conductivity due to magnons was reported for a canted AFM ordered phase in a noncentrosymmetric square-lattice AFM \cite{hotta19}.
As already mentioned in Ref.~\cite{buzo24}, it would be interested to work on the general conditions that yield a finite thermal Hall conductivity for a quantum paramagnetic phase.

%
In addition to a nearest-neighbor DM interaction, we also consider the effects of a next-nearest-neighbor DM one on a columnar VBS ground state. 
Indeed, previous studies, concerning the effects of a DM interaction 
in honeycomb lattice 
FMs \cite{owerre16,owerre16-2,wu21,kim22,Li26} and 
AFMs \cite{xiao16,kovalev16,hoon18,joshi19}, 
have considered a DM interaction between next-nearest-neighbor spins, with an out-of-plane DM vector pattern $\bD_{ij} = \pm D\hat{z}$, as illustrated in Fig.~\ref{fig:DM-vectors-nnn}:
In the absence of sublattice asymmetry between the triangular sublattices $A$ and $B$, a nearest-neighbor DM interaction is forbidden by the honeycomb lattice symmetries, while a next-nearest-neighbor one is allowed \cite{dm02}. 
Concerning the columnar VBS ground state (see Appendix \ref{appendix:DM-nnn} for details), we find that the effective boson model derived within the bond-operator formalism has quadratic and quartic terms (with respect to the number of triplet operators) only in the presence of sublattice asymmetry between the triangular sublattices $A$ and $B$.
In order to describe the columnar VBS phase within the harmonic approximation (see Sec.~\ref{sec:harmonic-approximation}), we then 
consider a next-nearest-neighbor DM interaction, with DM vectors
$\bD_{ij} = -D^A_z \, \hat{z}$, $\bD_{ij} = +D^B_z\,\hat{z}$, and $D^A_z \not= D^B_z$ (see Fig.~\ref{fig:DM-vectors-nnn}), similar to a previous study concerning a FM in a honeycomb lattice \cite{kim22}.   
In the presence of an external in-plane magnetic field $h_x$, we determine the energy of the triplon excitations within the harmonic approximation [see Fig.~\ref{fig:dispersion-nnn}(a)] and the Berry curvatures \eqref{eq:berry-curvature} of the triplon bands. Differently from the nearest-neighbor DM interaction \eqref{eq:d-directions}, a next-nearest-neighbor one does not yield finite Berry curvatures for the triplon bands.
Interestingly, in the presence of an out-of-plane magnetic field $h_z$, we find that the particle-hole symmetry \eqref{eq:diag-ham}, that is expected for the diagonal Hamiltonian, no longer holds. Apart from that, the energy of the triplon excitations can be calculated [see Fig.~\ref{fig:dispersion-nnn}(b)].
Again, the Berry curvatures of the triplon bands vanish throught the whole Brillouin zone, and therefore, both the Chern numbers of the triplon bands and the triplon thermal Hall response also vanish.
The physically more symmetric case $D_z^A=D_z^B$ requires a treatment of the cubic triplon interactions, lying beyond the harmonic approximation.

The thermal Hall response of honeycomb lattice AFMs in the absence of DM interactions have also been studied.
Fujiwara {\sl et al.} \cite{sota22} showed that an AFM in a honeycomb lattice might display a finite thermal Hall conductivity, once a spiral phase sets in and an on-site anisotropy is present. To exemplify such conditions, they considered spiral phases of the Heisenberg model \eqref{eq:ham-j1j2} on a honeycomb lattice and showed that, indeed, a finite thermal Hall conductivity is observed, regardless of the value of the external magnetic field. Interestingly, the Chern numbers of the magnon bands vanish for lower magnetic fields, but are finite ($C = \pm 1$) for higher ones (see Fig.~3 from Ref.~\cite{sota22} for details).
Mukherjee {\sl et al.} \cite{kundu23} studied the $J_1$-$J_2$ model \eqref{eq:j1j2} on a honeycomb lattice with an additional scalar (three-spin) chiral interaction $J_\chi$ within a Schwinger boson mean-field approach, similar to the ones employed in Refs.~\cite{zhang13,arnaud18}. It was found that, for lower chiral couplings $J_\chi$, 
a spin-liquid phase is stable for $0.24\, J_1 \lesssim J_2 \lesssim 0.36\, J_1$. 
For $J_2 \ge 0.32\, J_1$, a topological phase transition from a gapped spin-liquid (spinon bands with zero Chern numbers) to a gapped chiral one (spinon bands with nonzero Chern numbers) takes place at a critical chiral coupling.  
A finite thermal Hall conductivity due to spinons is found for the chiral spin-liquid phase 
(see Figs.~5 and 8 from Ref.~\cite{kundu23} for details).
It would be interesting to determine whether a scalar-chirality
interaction can stabilize a honeycomb-lattice columnar VBS with topologically
nontrivial triplon bands.

In summary, in this paper we studied the spin-$1/2$ $J_1$-$J_2$ AFM
Heisenberg model on a honeycomb lattice in the presence of 
inequivalent triangular sublattices, 
an additional nearest-neighbor DM interaction between the spins, and 
an external magnetic field. 
We considered the columnar VBS phase and studied the topological properties of the (elementary) triplon excitations. 
The VBS phase is described by an effective interacting boson model, determined within the bond-operator representation for spin operators.  
The analysis of the boson model within the harmonic approximation 
showed that the nearest-neighbor DM interaction provides a finite Berry curvature for the 
triplon bands, but the triplons are topologically trivial, since the Chern numbers of the triplon bands vanish. 
Interestingly, the thermal Hall conductivity due to triplons is finite at low temperatures.
Such features are similar to the ones found for a square-lattice columnar VBS phase \cite{buzo24}.

{\sl Note added.} Recently, we became aware of Ref.~\cite{sasamoto26}, which reports Schwinger boson mean-field theory results for the honeycomb-lattice $J_1$-$J_2$ model with an (out-of-plane) DM interaction between next-nearest-neighbor spins (see Fig.~\ref{fig:DM-vectors-nnn}) and an out-of-plane magnetic field. Three spin-liquid Ans\"atze were considered (the zero-, $\pi$-, and $\pi/2$-flux) and the corresponding spinon thermal Hall conductivity were determined. 
Interestingly, the $\kappa_{xy}$ of the zero-flux spin liquid state (see Fig.~4 from Ref.~\cite{sasamoto26}) is qualitatively similar to the one that we determined for the columnar VBS in the presence of a nearest-neighbor DM interaction (Fig.~\ref{fig:thermal-hall});
the increasing of the thermal Hall response with the DM interaction was also observed.
Our results though contrast with the ones of Ref.~\cite{sasamoto26}, since, for a next-nearest-neighbor DM interaction, we found a vanishing triplon thermal Hall response at the harmonic approximation when an asymmetry between the sublattices is considered (see Appendix~\ref{appendix:DM-nnn}).

\acknowledgments

We acknowledge financial support from the S\~ao Paulo Research Foundation (FAPESP), Brazil (Grant No. 2022/15453-0). E.M. acknowledges the financial support of CNPq, Brazil (Grant Nos. 309584/2021-3 and 309720/2026-5) and of the INCT project Advanced Quantum Materials, CNPq (Proc. 408766/2024-7), FAPESP (Proc. 2025/27091-3), and CAPES.
G.F.F. kindly acknowledges the financial support of the Coordena\c{c}\~ao de Aperfei\c{c}oamento de Pessoal de N\'ivel Superior - Brasil (CAPES) - Finance Code 001. 
R.L.D. dedicates this article to the memory of Mark O. Goerbig.

\section*{Data Availability}
The data that support the findings of this study are available from
the corresponding author upon reasonable request.

\appendix

\begin{widetext}

\section{Effective boson model in real space}
\label{appendix:effective-boson-model-in-real-space}

In this Appendix, we show the real-space expressions for the Hamiltonians in Eqs.~\eqref{eq:simp-ham-j1j2}, \eqref{eq:simp-ham-dm}, and \eqref{eq:dim-hamb}. The expressions are written in terms of the singlet $s_i$ and triplet $t_{i\, \alpha}$ operators 
($\alpha = x, y, z$), where $i$ is a site of the dimerized lattice $\mD$.
Further details can be found in the Appendix A from Ref.~\cite{buzo24}.

We first consider the $J_1$-$J_2$ model \eqref{eq:ham-j1j2}, whose expression in terms of the sites of the dimerized lattice $\mD$ is given by Eq.~\eqref{eq:dim-ham-j1j2}. Plugging the bond-operator representation \eqref{eq:bond-op-representation} into the Hamiltonian of Eq.~\eqref{eq:dim-ham-j1j2} and using the identity (A2) from Ref.~\cite{buzo24} for the local terms $J_1 (\bS^1_i \cdot \bS^2_i)$, the $\mH_{J, n}$ terms in Eq.~\eqref{eq:simp-ham-j1j2} are given by
\begin{align}
    \mH_{J,0} =& -\frac{3}{4} \Jone \sum_{i \in \mD} \sd_i s_i, 
    \nonumber  \\  
    \nonumber \\
    \mH_{J,2} =& \frac{\Jone}{4} \sum_{i \in \mD} \td_{i \alpha} t_{i \alpha} + \frac{1}{4} \sum_{i \in \mD, b} f_0^{\de{2}} \de{\bb}
             \de{ \sd_i \sd_{i + b} t_{i \alpha} t_{i + b, \alpha} + \sd_i s_{i + b} t_{i \alpha} \td_{i + b, \alpha} + \textrm{h.c.} },
    \nonumber \\  
    \nonumber \\
    \mH_{J,3} =& -\frac{i}{4} \sum_{i \in \mD, b} \epsilon_{\alpha \beta \gamma}  \Big[ \de{ f_0^{\de{3}}(\bb) + f_1^{\de{3}}(\bb) }  
             \de{ \sd_i t_{i \alpha} + \text{h.c.} } \td_{i + b, \beta} t_{i + b, \gamma}  
             - \de{ f_0^{\de{3}}(\bb) - f_1^{\de{3}}(\bb) }
              \td_{i \beta} t_{i \gamma} \de{ \sd_{i + b} t_{i + b, \alpha} + \text{h.c.} } \Big], 
    \nonumber \\  
    \nonumber \\
    \mH_{J,4} =& \frac{1}{4} \sum_{i \in \mD, b} f_0^{\de{4}} \de{\bb} \epsilon_{\alpha \beta \gamma} \epsilon_{\alpha \mu \nu} \td_{i \beta} t_{i \gamma} \td_{i + b, \mu} t_{i + b, \nu},
    \label{eq:real-hamj1j2}
\end{align}
where the summation convention over repeated greek indices is implied, the index $b$ indicates the dimer nearest-neighbor vectors $\bb_n$, with $n = 1,2,3$ [see Eq.~\eqref{eq:dim-vec}], and the auxiliary functions $f_m^{\de{n}} (\bb)$ read 
\begin{align}
    &f_0^{\de{2}} \de{\bb} =  -\Jone \de{\delta_{\bb, 2} + \delta_{\bb, -3}} 
    + \de{\Jtwo^A + \Jtwo^B} \de{ \delta_{\bb, 1} + \delta_{\bb, 2} + \delta_{\bb, 3} }, 
    \nonumber \\ \nonumber \\
    &f_0^{\de{3}} \de{\bb} = \Jone \de{ \delta_{\bb, 2} + \delta_{\bb, -3} }, 
    \nonumber \\ \nonumber \\
    &f_1^{\de{3}} \de{\bb} =  \de{\Jtwo^A - \Jtwo^B} \de{ \delta_{\bb, 1} + \delta_{\bb, 2} + \delta_{\bb, 3} },
    \nonumber \\ \nonumber \\
    &f_0^{\de{4}} \de{\bb} = \Jone \de{\delta_{\bb, 2} + \delta_{\bb, -3}} 
    + \de{\Jtwo^A + \Jtwo^B} \de{ \delta_{\bb, 1} + \delta_{\bb, 2} + \delta_{\bb, 3} }.
    \label{eq:aux-funcs1} 
\end{align}

Focusing on the DM interaction \eqref{eq:ham-dm}, the components $\mH_{DM, n}$ of Hamiltonian~\eqref{eq:simp-ham-dm} can be expressed as
\begin{align}
    \mH_{DM, 1} =& \frac{i}{2} \sum_{i \in \mD} \Big[ D_{\perp} \de{ \sd_i t_{i x} - \td_{i x} s_i } 
               + D_{\parallel} \de{ \sd_i t_{i y} - \td_{i y} s_i } \Big],
    \nonumber \\  
    \nonumber \\
    \mH_{DM, 2} =& \frac{1}{4} \sum_{i \in \mD, b} \de{ D_{\perp} g_\perp \de{\bb} + D_{\parallel} g_\parallel \de{\bb} }  
               \de{ \sd_{i} \sd_{i + b} t_{i \beta} t_{i + b, \gamma} + \sd_i s_{i + b} t_{i \beta} \td_{i + b, \gamma} + \textrm{h.c.} },
    \nonumber  \\  
    \nonumber \\
    \mH_{DM, 3} =& -\frac{i}{4} \sum_{i \in \mD, b} \de{ D_{\perp} g_\perp \de{\bb} + D_{\parallel} g_\parallel \de{\bb} } 
               \Big[ \epsilon_{\beta \mu \nu} \td_{i \mu} t_{i \nu} \de{ \sd_{i + b} t_{i + b, \gamma} + \textrm{h.c.} }
               - \epsilon_{\gamma \mu' \nu'} \de{ \sd_i t_{i \beta} + \textrm{h.c.} } \td_{i + b, \mu'} t_{i + b, \nu'} \Big],
    \nonumber \\  
    \nonumber \\
    \mH_{DM, 4} =& \frac{1}{4} \sum_{i \in \mD, b} \de{ D_{\perp} g_\perp \de{\bb} + D_{\parallel} g_\parallel \de{\bb} }
               \epsilon_{\beta \mu \nu} \epsilon_{\gamma \mu' \nu'} \td_{i \mu} t_{i \nu} \td_{i + b, \mu'} t_{i + b, \nu'},
    \label{eq:real-hamdm}
\end{align}
where the auxiliary functions $g_\perp\de{\bb}$ and $g_\parallel\de{\bb}$ are given by
\begin{align}
    g_\perp \de{\bb} =& \de{ 1/2 \epsilon_{x \beta \gamma} - \sqrt{3}/2 \epsilon_{y \beta \gamma} } \delta_{\bb, 2} 
                      + \de{ 1/2 \epsilon_{x \beta \gamma} + \sqrt{3}/2 \epsilon_{y \beta \gamma} } \delta_{\bb, -3},
    \nonumber \\
    \label{eq:aux-func-dm} \\
    g_\parallel \de{\bb} =& \de{ \sqrt{3}/2 \epsilon_{x \beta \gamma} + 1/2 \epsilon_{y \beta \gamma} } \delta_{\bb, 2} 
                      + \de{ -\sqrt{3}/2 \epsilon_{x \beta \gamma} + 1/2 \epsilon_{y \beta \gamma} } \delta_{\bb, -3}.
    \nonumber
\end{align}
Here, Eqs.~\eqref{eq:bond-op-representation} and \eqref{eq:dim-ham-dm} and the identity (A3) from Ref.~\cite{buzo24} were employed. 

Finally, once the generalized bond-operator representation~\eqref{eq:bond-op-representation} is inserted into the Zeeman term \eqref{eq:dim-hamb}, we find
\begin{equation}
    \label{eq:real-hamb}
    \mH_B = i \epsilon_{\alpha \beta \gamma} \sum_{i \in \mD} h_{\alpha} \td_{i \beta} t_{i \gamma},
\end{equation}
where $h_{\alpha} = g \mu_B B_{\alpha}$ (with $\alpha = x, \;y, \;z$).

\section{Dzyaloshinskii-Moriya interaction between next-nearest-neighbor spins: effective boson model for the columnar VBS phase}
\label{appendix:DM-nnn}

In this Appendix, we consider the columnar VBS phase in the honeycomb lattice in the presence of a DM interaction between next-nearest-neighbor spins. In this case, it can be shown \cite{dm02} that the DM interaction \eqref{eq:ham-dm} could have the out-of-plane DM vector pattern illustrated 
in Fig.~\ref{fig:DM-vectors-nnn}, even in the absence of asymmetry between sublattices $A$ and $B$.

\begin{figure}[t]
    \centering
    \includegraphics[width=0.3\linewidth]{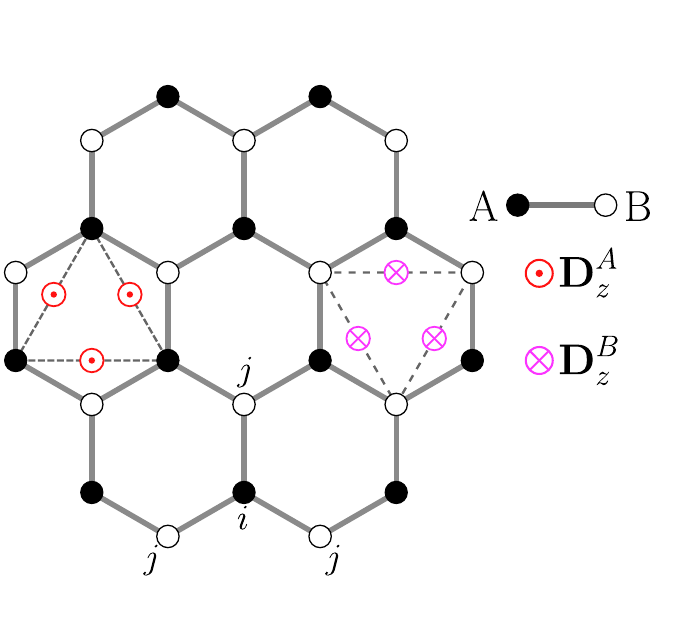}
    \caption{\justifying{Schematic representations of the DM vectors $\bD_{ij}$, allowed by the symmetries of the honeycomb lattice, for a DM interaction \eqref{eq:ham-dm} between next-nearest-neighbor spins,
    see Eqs.~\eqref{eq:dim-ham-dm-nnn} and \eqref{eq:DM-vectors-nnn} for details.
    } }
    \label{fig:DM-vectors-nnn}
\end{figure}

Expressed in terms of the dimerized lattice $\mD$, the DM interaction \eqref{eq:ham-dm} assumes the form 
\begin{align}
    \label{eq:dim-ham-dm-nnn}
    \mH_{DM} = \sum_{i \in \mD} \Big[&\bD^{A}_{i, i+1} \cdot \de{ \bS^1_i \times \bS^1_{i+1} } + \bD^{B}_{i, i+1} \cdot \de{ \bS^2_i \times \bS^2_{i+1} }
    + \bD^{A}_{i, i+2} \cdot \de{ \bS^1_i \times \bS^1_{i+2} } + \bD^{B}_{i, i+2} \cdot \de{ \bS^2_i \times \bS^2_{i+2} }
    \nonumber \\
    &+ \bD^{A}_{i, i+3} \cdot \de{ \bS^1_i \times \bS^1_{i+3} } + \bD^{B}_{i, i+3} \cdot \de{ \bS^2_i \times \bS^2_{i+3} } \Big],
\end{align}
where the index $n=1,\,2,\,3$ represents the dimer nearest-neighbor vectors \eqref{eq:dim-vec} and the DM vectors $\bD^{A,B}_{i, i+n}$ are defined as
\begin{equation}
    \bD^A_{i,i+n} = D^A_{z} \de{0,\;0,\;1}, 
    \quad\quad\quad 
    \bD^B_{i,i+n} = D^B_{z} \de{0,\;0,\;-1} 
    \label{eq:DM-vectors-nnn}.
\end{equation}
With the aid of the generalized bond-operator representation \eqref{eq:bond-op-representation}, 
the Hamiltonian \eqref{eq:dim-ham-dm-nnn} can be expressed as
\begin{equation}
    \label{eq:simp-ham-dm-nnn}
    \mH_{DM} = \mH_{DM,2} + \mH_{DM, 3} + \mH_{DM, 4},
\end{equation}
where the $\mH_{DM, n}$ terms, defined as in Eq.~\eqref{eq:simp-ham-dm}, are given by
\begin{align}
    \mH_{DM,2} =& \sum_{i \in \mD, b} g^-_z(\bb) \epsilon_{z \beta \gamma} 
                \de{\sd_i \sd_{i + b} t_{i \beta} t_{i + b, \gamma} + \sd_{i} s_{i + b} t_{i \beta} \td_{i + b, \gamma} + \textrm{h.c.}}, 
    \nonumber \\  
    \nonumber \\
    \mH_{DM,3} =& -i \sum_{i \in \mD, b} g^+_z(\bb) \epsilon_{z \beta \gamma}
                \Big[\epsilon_{\beta \mu \nu} \td_{i \mu} t_{i \nu} \de{ \sd_{i + b} t_{i + b, \gamma} + s_{i + b} \td_{i + b, \gamma} }
                + \epsilon_{\gamma \mu' \nu'} \de{ \sd_i t_{i \beta} + s_{i} \td_{i \beta} } \td_{i + b, \mu'} t_{i + b, \nu'} \Big],
    \label{eq:real-hamdm-nnn} \\
    \nonumber \\
    \mH_{DM,4} =& -\sum_{i \in \mD, b} g^-_z(\bb) \epsilon_{z \beta \gamma} \epsilon_{\beta \mu \nu} \epsilon_{\gamma \mu' \nu'} 
                \de{ \td_{i \mu} t_{i \nu} \td_{i + b, \mu'} t_{i + b, \nu'} },
    \nonumber   
\end{align}
with $\epsilon_{\alpha \beta \gamma}$ being the totally antisymmetric tensor ($\epsilon_{xyz} = 1$) 
and $g^\pm_z$ being the auxiliary functions
\begin{equation}
    \label{eq:aux-func-dm-nnn}
    g^\pm_z(\bb) = \frac{1}{4} \de{ D^A_{z} \pm D^B_{z} } \de{ \delta_{\bb, \bb_1} + \delta_{\bb, \bb_2} + \delta_{\bb, \bb_3} }.
\end{equation}
One notes that, for equivalent sublattices, $D^A_z = D^B_z$, which yields $g^-_z (\bb) = 0$, and therefore, the quadratic $\mH_{DM,2}$ and quartic $\mH_{DM,4}$ terms vanish.

In order to capture the effects of the DM interaction between next-nearest-neighbor spins at the harmonic approximation, as done in the main text, we here also consider finite sublattice asymmetry, i.e., we assume that $D_z^A \neq D_z^B$.
Fourier transforming the triplet operators $t_{i\alpha}$ [see Eq.~\eqref{eq:fourier-transform}], 
we obtain the momentum-space representation of the Hamiltonian $\mH_{DM, 2}$,
\begin{equation}
    \mH_{DM, 2} = i \sum_{\bk} E_{\bk} \epsilon_{z \beta \gamma} \Big[ \td_{\bk \beta} t_{\bk \gamma} + \frac{1}{2} \de{t_{-\bk, \beta} t_{\bk \gamma} - \td_{\bk \gamma} \td_{-\bk \beta}}\Big],
    \label{eq:hdm2-k-nnn}
\end{equation}
where the coefficient $E_{\bk}$ reads
\begin{equation}
    \label{eq:Ek-coef}
    E_{\bk} = \frac{N_0}{2} \de{D_z^A - D_z^B} \de{\sin{k_1} + \sin{k_2} + \sin{k_3}},
\end{equation}
with $k_n$ ($n = 1,2,3$) being given by Eq.~\eqref{eq:cdot-kb}.
Including the terms up to quadratic order in the triplet operators derived from the $J_1$-$J_2$ model [Eqs.~\eqref{eq:fourier-ham0-j1j2} and \eqref{eq:fourier-ham2-j1j2}] and the Zeeman term \eqref{eq:fourier-hamb}, 
we find that the total Hamiltonian within the harmonic approximation also assumes the form \eqref{eq:harmonic-approx}, 
with the $6\times6$ matrix $\Hk$ and the $3\times 3$ matrix $\HB$ being given by Eqs.~\eqref{eq:Hk} and \eqref{eq:HB}, respectively, but the $3\times3$ matrices $\Ak$ and $\Bk$ now read
\begin{equation}
    \Ak = 
    \begin{pmatrix}
        A_{\bk}   & iE_{\bk} & 0 \\
        -iE_{\bk} & A_{\bk}  & 0 \\
        0         & 0        & A_{\bk}
    \end{pmatrix},
\quad\quad\quad
    \Bk = 
    \begin{pmatrix}
        B_{\bk}   & iE_{\bk} & 0 \\
        -iE_{\bk} & B_{\bk}  & 0 \\
        0         & 0        & B_{\bk}
    \end{pmatrix};
\label{eq:Ak-nnn}    
\end{equation}
here, the coefficients $A_\bk$ and $B_\bk$ are also given by Eq.~\eqref{eq:coefs-hj}. 
In the following, we consider the effects of an external in-plane magnetic field (perpendicular to the $\hat{z}$ axis) and the ones of an out-of-plane magnetic field (along the $\hat{z}$ axis); in both cases, we assume that
\begin{equation}
    \label{eq:dza-dzb-coupling}
    D_z^B = \alpha D_z^A = \alpha D_z,
\end{equation}
with a fixed $\alpha = 0.90$. 

\begin{figure}
    \centering
    \begin{subfigure}[b]{0.40\textwidth}
        \centering
        \llap{\textbf{(a)}\hspace{-0.05em}}
        \vspace*{-1.5em}
        \includegraphics[width=\textwidth]{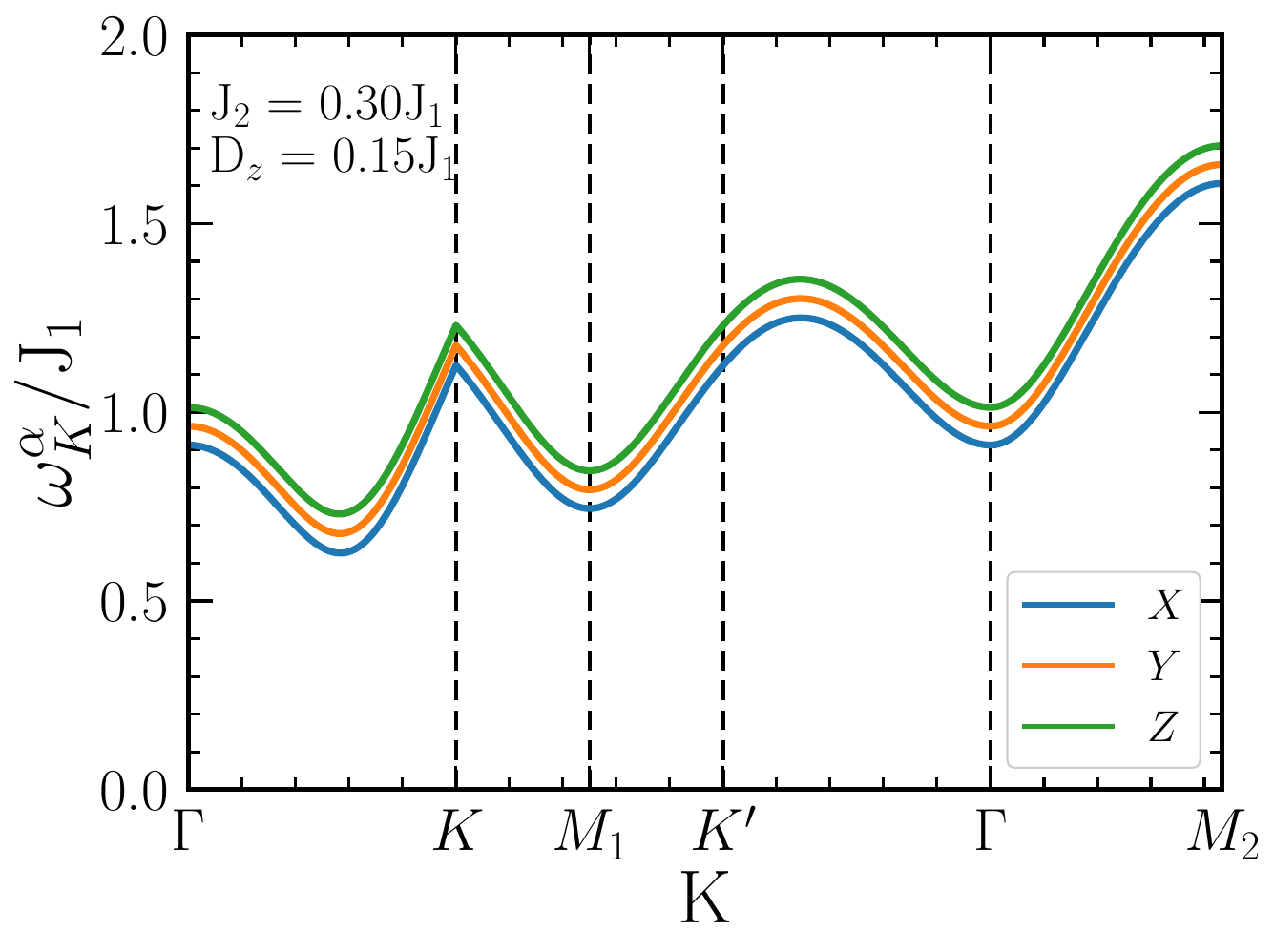}
    \end{subfigure}
    \hspace{1.7cm}
    \begin{subfigure}[b]{0.40\textwidth}
        \centering
        \llap{\textbf{(b)}\hspace{-0.05em}}
        \vspace*{-1.5em}
        \includegraphics[width=\textwidth]{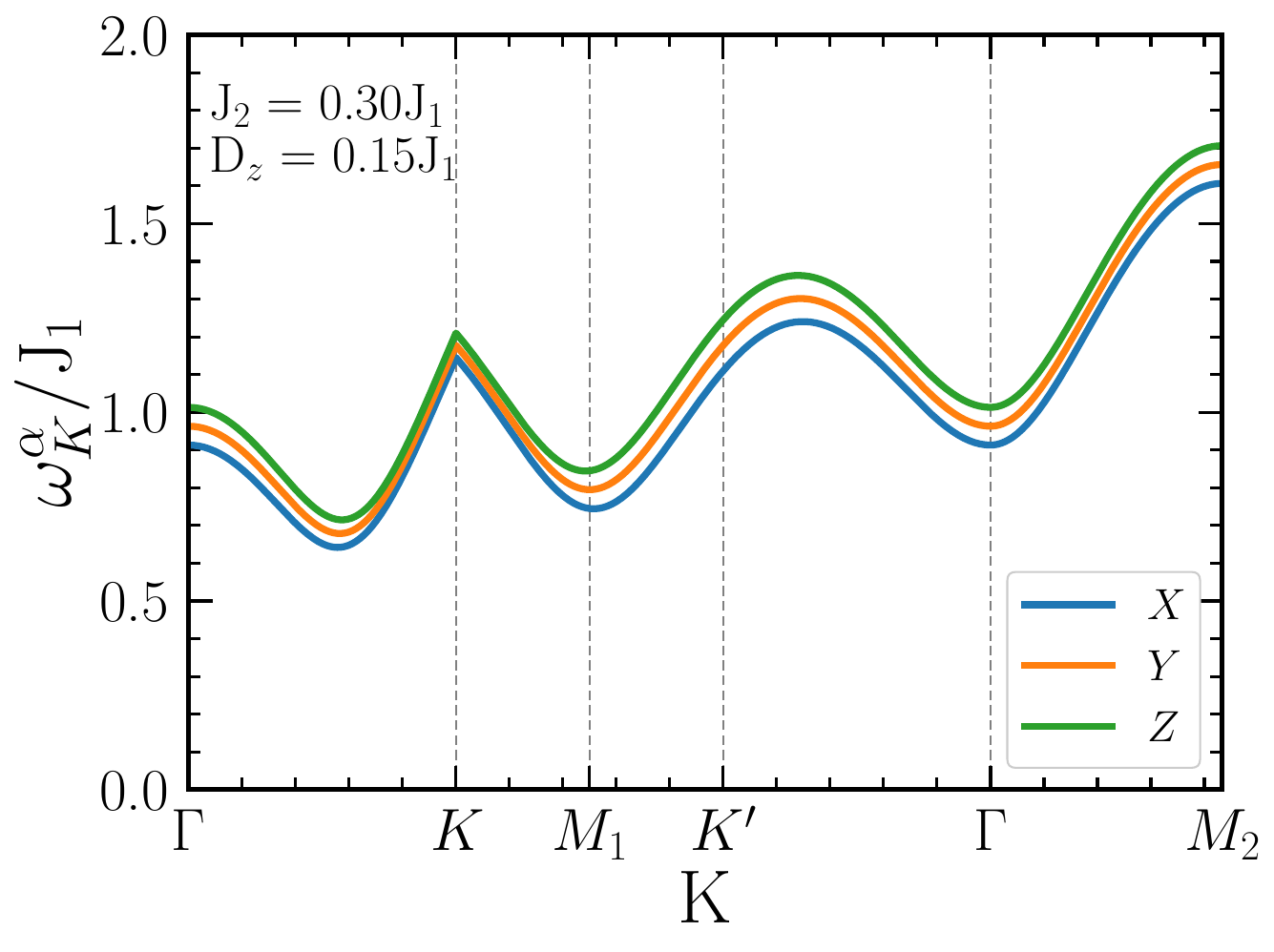}
    \end{subfigure}
    \caption{\justifying{Triplon dispersion relations 
    $\wk^x$ (solid blue line),  $\wk^y$ (solid orange line), and  $\wk^z$ (solid green line) 
    along a path in the dimerized Brillouin zone 
    for the columnar VBS phase of the Heisenberg model \eqref{eq:ham}, considering the next-nearest-neighbor DM interaction \eqref{eq:ham-dm} characterized by the DM vectors \eqref{eq:DM-vectors-nnn}.  
    Results derived within the harmonic approximation (see Sec.~\ref{sec:harmonic-approximation}).
    (a) $J_2 = 0.30 \, J_1$, $D_z = 0.15 \, J_1$, and $h_x = -0.05 \, J_1$; 
    (b) $J_2 = 0.30 \, J_1$, $D_z = 0.15 \, J_1$, and $h_z = -0.05 \, J_1$.  }}
    \label{fig:dispersion-nnn}
\end{figure}

Let us consider an external in-plane magnetic field $h_x = -0.05\,J_1$. 
Here, the condition \eqref{eq:colpa-condition} is satisfied, and therefore the diagonalization procedure outlined in Sec.~\ref{sec:harmonic-approximation} can be employed.
By numerically solving the self-consistent equations \eqref{eq:self-const}, we calculate the parameters $N_0$ and $\mu$,
the ground-state energy $E_{GS}$, the triplon excitation gap $\Delta$, and the momentum $\bQ$ associated with the triplon gap. 
We find that the dependence of such quantities with the next-nearest-neighbor DM interaction $D_z$ is qualitatively similar to the 
ones shown in Figs.~\ref{fig:params1} and \ref{fig:params2}, and therefore these plots are omitted.
Again, we observe that the DM interaction $D_z$ alone induces a partial gap opening in the triplon dispersion relation, a feature also displayed by the nearest-neighbor DM interaction $D_\parallel$, see Figs.~\ref{fig:dispersion}(a) and \ref{fig:dispersion}(b);
as exemplified in Fig.~\ref{fig:dispersion-nnn}(a), the degeneracies of the triplon dispersion relation are completely lifted in the presence of an external magnetic field, see also Figs.~\ref{fig:dispersion}(c) and \ref{fig:dispersion}(d).  
Finally, the Berry curvatures $\Omega_\alpha(\bk)$ of the triplon bands [Eq.~\eqref{eq:berry-curvature}] identically vanish across the entire hexagonal Brillouin zone; such a behaviour stands in stark contrast to the pronounced and high-intensity peaks shown in Fig.~\ref{fig:berry-curvature} for the Berry curvatures of the triplon bands in the presence of a nearest-neighbor DM interaction $D_\parallel$. As a direct consequence, both the Chern numbers $C_\alpha$ of the triplon bands [Eq.~\eqref{eq:chern-number}] 
and the thermal Hall conductivity $\kappa_{xy}$ due to triplons [Eq.~\eqref{eq:thermal-hall}] are strictly zero;
for the moment, the underlying physical mechanism responsible for such exact cancellation of the Berry curvatures under a next-nearest-neighbor DM coupling $D_z$ and an in-plane magnetic field $h_x$ remains an open question.

For an applied out-of-plane magnetic field $h_z$, the corresponding $6\times6$ matrix $\Hk$ again obeys the condition \eqref{eq:colpa-condition}, but the resulting diagonal matrix $\Hdiag$ does not have the particle-hole symmetry \eqref{eq:Hkdiag} previously found above for an in-plane magnetic field and in Sec.~\ref{sec:harmonic-approximation}, see Eq.~\eqref{eq:Hkdiag}. 
Here, we indeed find that
\begin{equation}
    \label{eq:Hkdiag2}
    \Hdiag = \textrm{diag}\de{ \wk^{x},\; \wk^{y},\; \wk^{z},\; 
             \bar{\omega}_\bk^{x},\; \bar{\omega}_\bk^{y}, \; \bar{\omega}_\bk^{z} },
\end{equation}
with $\wk^\alpha \not= \bar{\omega}_\bk^\alpha$. 
In particular, for the DM interaction defined by the DM vectors \eqref{eq:DM-vectors-nnn}, we have
$\bar{\omega}_\bk^\alpha = \omega_{-\bk}^\alpha$. 
Apart from that issue, the dispersion relation of the triplon excitations can be determined, as shown, e.g., in Fig.~\ref{fig:dispersion-nnn}(b).   
Similarly to the in-plane magnetic field case discussed above, here the Berry curvatures \eqref{eq:berry-curvature} of the triplon bands also vanish, 
and therefore, the Chern numbers $C_\alpha$ of the triplon bands  
and the triplon thermal Hall response $\kappa_{xy}$ are also zero.

\end{widetext}

\bibliography{refs}


\end{document}